\documentclass{jfm}
\usepackage{graphicx}
\usepackage{newtxtext}
\usepackage{newtxmath}
\usepackage{natbib}
\usepackage{bm}
\usepackage{hyperref}
\usepackage{ upgreek }
\usepackage{multirow}
\usepackage[dvipsnames]{xcolor}
\hypersetup{
    colorlinks = true,
    urlcolor   = blue,
    citecolor  = black,
}

\newcommand{\RomanNumeralCaps}[1]
\linenumbers

\title{Contrary to Newtonian trends: Early flow transition and drag enhancement at low to intermediate Reynolds number flows of structured fluids}

\author{Kartik\aff{1}, Anant Chauhan\aff{1},
 Chandi Sasmal\aff{1}
 \corresp{\email{csasmal@iitrpr.ac.in}}}

\affiliation{\aff{1}Department of Chemical Engineering, Indian Institute of Technology Ropar, Rupnagar, Punjab, India-140001}

\begin{document}
\maketitle

\begin{abstract}
The flow of structured fluids, such as polymeric and micellar solutions, involves a strong two-way coupling between flow kinematics and internal microstructural dynamics, including polymer stretching, micellar scission, and reformation. These interactions yield complex nonlinear rheological responses, including viscoelasticity, shear-thinning, and thixotropy. In this study, we perform high-fidelity numerical simulations of micellar solution flow past a circular cylinder using the Modified Bautista-Manero (MBM) model, which couples a viscoelastic constitutive equation with a kinetic equation for fluidity to capture reversible micellar breakage and reformation. Model parameters are derived from quantitative fitting of experimental data for the EHAC-NaSal system. Our results reveal substantially richer flow dynamics than in Newtonian fluids under the same conditions. Transitions to unsteady flow occur at significantly lower Reynolds numbers, indicating greater instability driven by microstructural effects. At intermediate Reynolds numbers, micellar flows exhibit quasi-periodic behaviour, contrasting with classical periodic vortex shedding in Newtonian cases. Unlike the monotonic drag reduction in Newtonian fluids, micellar solutions show an anomalous drag increase. Wake characteristics reverse with regime: larger recirculation zones at low Reynolds numbers but more compact wakes in unsteady flows. Lift coefficients and Strouhal numbers are consistently increased. Vorticity fields display pronounced spatial localisation within thin near-wake shear layers. A dynamic mode decomposition analysis reveals coexisting unstable time-decaying and self-sustained modes in micellar flows, whereas only self-sustained modes appear in Newtonian cases. These findings highlight the critical role of microstructural kinetics in dictating flow transitions, vortex dynamics, and force characteristics, with direct implications for applications involving surfactant-based fluids.   
\end{abstract}


\section{Introduction\label{sec:intro}}
Structured fluids are a broad class of complex fluids characterised by an internal microstructure arising from the spatial organisation of their constituents, such as polymer molecules, surfactant assemblies, colloidal particles, droplets, or biological entities~\citep{witten2004structured}. Their common varieties include polymer solutions and melts, wormlike micellar solutions, colloidal suspensions, emulsions, gels, foams, and biological fluids such as blood. Owing to their evolving microstructure, structured fluids exhibit a range of complex flow behaviours, including shear-thinning or shear-thickening, viscoelasticity, yield stress, and thixotropy, which are absent in simple Newtonian fluids~\citep{de1994rheological}. These unique rheological characteristics make structured fluids central to numerous applications spanning food processing, cosmetics, pharmaceuticals, coatings, enhanced oil recovery, biomedical devices, drug delivery systems, and emerging technologies such as soft robotics and additive manufacturing, where precise control of flow and mechanical response is essential~\citep{ghaffarkhah2024chemistry}.

The complex rheological behaviour of structured fluids arises from the continuous interplay between microstructural breakdown and reconstruction under applied deformation. A prototypical example is provided by wormlike micellar solutions (WLMs), which are self-assembled surfactant systems composed of elongated, flexible micelles that form transient, entangled networks and undergo reversible scission and recombination during flow~\citep{chu2013smart,yang2002viscoelastic}. This dynamic and reversible microstructural evolution imparts pronounced time- and history-dependent material responses, giving rise to multiple, often strongly nonlinear rheological behaviours, including viscoelasticity, thixotropy, and viscoplasticity~\citep{walker2001rheology,wheeler1996structure}. As a consequence, WLMs exhibit complex flow phenomena even in the creeping-flow regime, where inertial effects are negligible compared to viscous and elastic stresses. In particular, numerous experimental studies have demonstrated that WLM flows become elastically unstable once the flow strength, typically quantified by the Weissenberg number, exceeds a critical threshold. With a further increase in this dimensionless parameter, the flow transitions from steady, symmetric states to asymmetric states and eventually to time-dependent, chaotic dynamics, commonly referred to as elastic turbulence~\citep{zhao2014microfluidic}. For example, Dubash et al.~\citep{dubash2012elastic} investigated WLM flow in a microfluidic cross-slot geometry using a cetyltrimethylammonium bromide (CTAB)-sodium salicylate (NaSal) system and reported a sequence of transitions from steady symmetric to steady asymmetric and ultimately unsteady flow as the Weissenberg number increased beyond a critical value. Similar flow transitions were observed by Haward et al.~\citep{haward2012extensional} for the same micellar system, while subsequent work demonstrated that the onset and nature of these instabilities are strongly influenced by surfactant concentration and ionic strength~\citep{haward2012stagnation}. Beyond cross-slot configurations, comparable elastic instabilities and flow transitions have been reported in other low-Reynolds-number flow systems involving WLMs, including flow past cylindrical micropillars in microchannels, where increasing flow strength similarly drives transitions from steady to unsteady flow regimes~\citep{zhao2016flow,haward2019flow,hopkins2022effect,moss2010flow}.

In addition to extensive experimental investigations, a substantial body of numerical work has been devoted to elucidating the flow behaviour of wormlike micellar solutions across a wide range of flow configurations, offering enhanced mechanistic insight into experimentally observed phenomena. For example, Kalb et al.~\citep{kalb2017role,kalb2018elastic} demonstrated through numerical simulations of microfluidic cross-slot flows that micellar chain scission plays a pivotal role in governing the onset and development of elastic instabilities. Their study employed the Vasquez-Cook-McKinley (VCM) constitutive model, which explicitly incorporates the coupled kinetics of micellar breakage and recombination via a structural evolution equation~\citep{vasquez2007network}. Using the same modelling framework, Khan and Sasmal~\citep{khan2020effect,khan2021elastic} further showed that flow-induced micellar scission significantly alters the instability characteristics in flows past a microfluidic cylinder, highlighting the strong dependence of elastic instabilities on microstructural dynamics. Similar elastic instabilities have been reported in numerical and experimental studies of WLM flows through other complex geometries, including microporous media~\citep{sasmal2020flow,khan2022effect}, Taylor-Couette systems~\citep{rossi2006slippage,mohammadigoushki2017inertio,ghadai2023origin}, and flow past spherical particles~\citep{chen2004flow,mohammadigoushki2016sedimentation,jayaraman2003oscillations,sasmal2021unsteady}. Comprehensive overviews of these rich and multiscale flow behaviors across different geometries are available in several review articles~\citep{rothstein2020complex,fardin2012instabilities}. Beyond the creeping-flow regime, comparatively fewer studies have explored wormlike micellar flows in the high-turbulence regime, despite their well-established drag-reducing capabilities. Notably, Fukushima et al.~\citep{fukushima2022modification} experimentally investigated two-dimensional turbulent flow past an array of circular cylinders and demonstrated that flow-induced micellar structures exert a strong influence on vortex shedding characteristics and turbulence statistics. A number of studies focusing on turbulent drag reduction in wormlike micellar solutions attribute these effects to the formation, evolution, and eventual destruction of shear-induced structures (SIS) within the flow field~\citep{anh2013advection,wakimoto2018simultaneous,mitishita2022turbulent,ohlendorf1986surfactant}.

Therefore, the existing body of literature indicates that the flow behaviour of wormlike micellar solutions has been extensively studied in the creeping-flow regime and, to a lesser extent, in turbulent conditions. However, the low-to-intermediate Reynolds number regime remains largely unexplored, even for canonical configurations such as flow past a circular cylinder. In contrast, for Newtonian fluids, both experimental and numerical investigations have comprehensively established the sequence of flow transitions with increasing Reynolds number, ranging from steady, symmetric flow without separation to steady recirculating wakes, then to periodic vortex shedding, and ultimately to chaotic turbulent states~\citep{williamson1996k,zdravkovich1997flow}. Similar studies have also been extended to inelastic generalised Newtonians~\citep{chhabra2011fluid} and viscoelastic fluids~\citep{hamid2022dynamic,oliveira2001method,richter2010simulations}, revealing that shear-thinning behaviour typically delays the onset of unsteadiness, while shear-thickening promotes earlier transition~\citep{bharti2006steady,sivakumar2006effect,d1996steady,patnana2009two}. Additionally, viscoelastic effects are known to postpone boundary-layer separation and vortex shedding, significantly altering wake dynamics and force coefficients~\citep{hamid2022dynamic,sahin2004effects,usui1980karman}.

Despite these advances, a critical gap persists in understanding how such flow transitions manifest in wormlike micellar solutions, where shear-thinning, viscoelasticity, and thixotropy coexist due to continuous microstructural breakdown and reformation~\citep{walker2001rheology,dreiss2007wormlike}. Addressing this gap requires a constitutive framework capable of capturing the coupled evolution of stress and microstructure. In this study, the Modified-Bautista-Manero (MBM) model is employed, as it incorporates a structure-dependent fluidity parameter that is governed by competing breakdown and rebuilding kinetics~\citep{boek2005constitutive}, enabling a unified description of time-dependent viscosity, elasticity, and structural memory effects. Furthermore, a data-driven Dynamic Mode Decomposition (DMD) approach is also utilised to extract dominant coherent structures, growth rates, and characteristic frequencies directly from the flow field~\citep{schmid2011}, providing quantitative insight into how microstructural dynamics modify vortex shedding behaviour relative to that in Newtonian flows under otherwise identical conditions. In addition, detailed assessments of global and local flow characteristics, such as drag and lift coefficients, are conducted over a range of Reynolds numbers. Overall, the findings offer new fundamental insights into vortex-microstructure interactions and provide a basis for leveraging surfactant-based additives to control wake dynamics and hydrodynamic forces in practical applications.

\section{Problem description and governing equations\label{sec:probdes}}

\begin{figure}
    \centering
    \includegraphics[width=13cm]{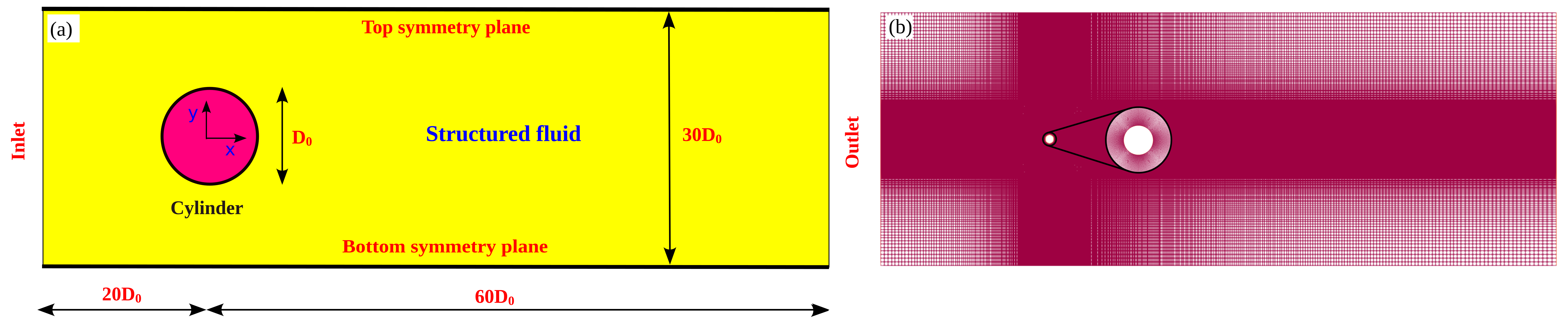}
    \caption{(a) Schematic of the present problem. Here, $D_0$ is the cylinder diameter, and all the computational details, such as different boundaries and distances, are properly labelled in the schematic. (b) Grid structure G-3 with refined cells near the cylinder surface, shown in the zoom-in to capture the steep gradients of velocity, pressure, stress, fluidity, etc.}
    \label{schematicAndGrid}
\end{figure}

As mentioned in the preceding section, the problem considered herein is the study of the flow dynamics of structured fluids, particularly wormlike micellar solutions, past a transversely placed circular cylinder of diameter $D_{0}$, as schematically shown in figure~\ref{schematicAndGrid}. To make the present flow system computationally tractable, a rectangular fluid domain of length 80$D_{0}$ and of height 30$D_{0}$ is chosen in this study. The cylinder is placed 20$D_{0}$ from the inlet and at the centre of this fluid domain~\citep{hamid2022dynamic}. The structured fluid enters the computational domain with a uniform velocity of $U_{\infty}$ through the inlet section. Assuming the fluid is to be incompressible, the following equations will govern the present flow field, written in their dimensional forms\\
\newline
Continuity equation\\  
\begin{equation}
    \bm{\nabla} \cdot \bm{u} = 0
\end{equation}
Cauchy's momentum equation\\
\begin{equation}
    \rho \left( \frac{\partial \bm{u}}{\partial t} + \bm{u} \cdot \bm{\nabla} \bm{u}  \right) = -\bm{\nabla} p + \bm{\nabla} \cdot \bm{\tau}
\end{equation}
In the above equations, $\rho$ is the fluid density, $\bm{u}$ is the velocity vector, $t$ is the time, $p$ is the pressure, and $\bm{\tau}$ is the extra or deviatoric stress tensor. For evaluating the last term, the Modified-Bautista-Manero (MBM) constitutive model is used~\citep{boek2005constitutive}, as detailed below.

\subsection{Modified-Bautista-Manero (MBM) constitutive model for wormlike micellar solutions}
The original BMP model~\citep{bautista1999understanding} consists of a set of equations, namely, the upper convected Maxwell (UCM) constitutive equation, which calculates the stress contribution from the fluid microstructure, such as the micellar network, in the present case, and a kinematic equation proposed by Fredrickson~\citep{fredrickson1970model}, which introduces a structural parameter for describing molecular entanglement, network junctions and destruction of structures. The following equations govern these two mechanisms

\begin{equation}
    \bm{\tau}^* + \frac{1}{G_{0}\phi^*} \overset{\nabla}{\bm{\tau}^*} = \frac{\dot{\bm{\gamma}}^*}{\phi^*}
\end{equation}

\begin{equation}\label{bmp}
    \frac{\partial \phi^*}{\partial t^*} + \bm{u}^* \cdot \bm{\nabla}^* \phi^* = \frac{1}{\lambda} (\phi_{0} - \phi^*) + K_{0} (\phi_{\infty} - \phi^*) \bm{\tau}^*: \dot{\bm{\gamma}^*}
\end{equation}

In the above equations, $\overset{\nabla}{\bm{\tau^*}}$ is the upper convected derivative of the stress tensor defined as $\overset{\nabla}{\bm{\tau^*}} = \frac{\partial \bm{\tau^*}}{\partial t^*} + \bm{u}^* \cdot \bm{\nabla}^* \bm{u}^* - (\bm{\nabla}^* \bm{u}^*)^{T} \cdot \bm{\tau}^* - \bm{\tau}^* \cdot (\bm{\nabla}^* \bm{u}^*)$, $G_{0}$ is the relaxation modulus, and $\phi^*$ is the structural parameter called fluidity (defined as the inverse of the viscosity, i.e., $\phi^* \equiv \eta^{-1}$). The first term in the right-hand side of equation~\ref{bmp} represents the reformation process caused by the build-up of viscosity (or the breakdown of fluidity), where $\lambda$ is the structural relaxation time, $\phi_{0} (\equiv \eta_{0}^{-1})$ is the plateau fluidity observed at low shear rates representing the fully structured state of the fluid microstructure. The second term represents the destruction process of the fluid microstructure, resulting in the breakdown of viscosity (or the build-up of fluidity), where $K_{0}$ is a parameter related to structure breakdown (with unit of $\text{Pa}^{-1}$), $\phi_{\infty} (\equiv \eta_{\infty}^{-1})$ is the fluidity at high shear rates representing the completely unstructured state of the fluid microstructure, and $\bm{\tau}^*: \dot{\bm{\gamma}^*}$ represents the rate of energy dissipation because of the fluid microstructure breakdown process. Here, $^*$ denotes a dimensional variable. 

One disadvantage of the original BMP model is that it predicts an unbounded extensional viscosity even at finite deformation rates, as demonstrated by Boek et al.~\citep{boek2005constitutive}. Therefore, they proposed a modification of the original BMP model by splitting the contribution of the extra stress tensor $\bm{\tau}^*$ into two components, namely, the contribution from the solvent $\bm{\tau_{s}}$ and from the fluid microstructure $\bm{\tau_{p}}$, i.e., $\bm{\tau}^* = \bm{\tau_{s}}^* + \bm{\tau_{p}}^*$, and the resulting model is called the Modified-Bautista-Manero constitutive model. According to this model, the following set of equations will govern the flow

\begin{equation}
    \bm{\tau_{s}}^* = \eta_{s} \bm{\dot{\gamma}}^*
\end{equation}

\begin{equation}
    \bm{\tau_{p}}^* + \frac{1}{G_{0}\phi^*} \overset{\nabla}{\bm{\tau_{p}^*}} = \frac{\dot{\bm{\gamma}^*}}{\phi^*}
\end{equation}

\begin{equation}\label{mbm}
    \frac{\partial \phi^*}{\partial t^*} + \bm{u}^* \cdot \bm{\nabla}^* \phi^* = \frac{1}{\lambda} (\phi_{0} - \phi^*) + \left(\frac{K_{0}}{\eta_{\infty}}\right) \bm{\tau_{p}^*}: \dot{\bm{\gamma}^*}
\end{equation}

In the above equations, $\frac{K_{0}}{\eta_{\infty}} = K_{0} \phi_{\infty}$ is considered as a single finite parameter in this modified BMP model. Furthermore, in the original BMP model, $\phi_{0} = \eta_{0}^{-1}$, where the zero-shear rate viscosity $\eta_{0} = (\eta_{s} + {\eta_{p}}_{0})$. In contrast, in the MBM model, $\phi_{0} = {\eta_{p}}_{0}^{-1}$. Finally, the above governing equations are non-dimensionalised using the following scaling variables: fluidity is scaled with $\phi_{0}$, length is scaled with $D_{0}$, velocity is scaled with $U_{\infty}$, time is scaled with $\frac{D_0}{U_{\infty}}$, fluid microstructure stresses and pressure are scaled with $\rho U_{\infty}^{2}$. The non-dimensional forms of the governing equations become as follows

\begin{equation}
    \bm{\nabla}\cdot \bm{u} = 0
\end{equation}

\begin{equation}
    \frac{\partial \bm{u}}{\partial t} + \bm{u} \cdot \bm{\nabla} \bm{u} = -\bm{\nabla} p + \frac{\beta}{Re} \bm{\nabla}^{2} \bm{u} + \bm{\nabla} \cdot \bm{\tau_{p}}
\end{equation}

%\begin{equation}
 %   \bm{\tau}_{s} = \frac{\beta}{Re} \dot{\bm{\gamma}}
%\end{equation}

\begin{equation}
    \bm{\tau_{p}} + \frac{Wi}{\phi} \overset{\nabla}{\bm{\tau_{p}}} = \frac{(1 - \beta)}{\phi~Re} \dot{\bm{\gamma}}
\end{equation}

\begin{equation}
    \frac{\partial \phi}{\partial t} + \bm{u} \cdot \bm{\nabla} \phi = \frac{1}{\Lambda} (1 - \phi) + \Gamma_{M}~Re~\bm{\tau_{p}} : \dot{\bm{\gamma}} 
\end{equation}

It can be seen that these governing equations are associated with five dimensionless numbers, namely, Reynolds number $Re \left( = \frac{D_{0}U_{\infty} \rho}{\eta_{0}} \right)$, Weissenberg number $Wi \left(= \lambda_{ve}\frac{U_{\infty}}{D_{0}} = \frac{1}{G_{0}\phi_{0}} \frac{U_{\infty}}{D_{0}}\right)$, construction number or Mnemosyne number or thixoviscous number $\Lambda \left( = \lambda \frac{U_{\infty}}{D_{0}} \right)$, destruction number $\Gamma_{M} \left(= \frac{K_{0}}{\eta_{\infty}} {\eta_{p}}_{0}(\eta_{s}+{\eta_{p}}_{0})\frac{U_{\infty}}{D_{0}} \right)$, and viscosity ratio $\beta \left( =\frac{\eta_{s}}{\eta_{s} + {\eta_{P}}_{0}} \right)$. Here, the thixoviscous number is associated with the reformation of complex fluid microstructures. In particular, it is a non-dimensional time scale over which fluid microstructure recovers from flow. A high value of $\Lambda$ (i.e., $\Lambda \rightarrow \infty$) signifies that the fluid will exhibit a slow or null recovery of fluid microstructure (or viscosity) after the cessation of flow, whereas a fast recovery will happen for the opposite case of a low value of $\Lambda$ (i.e., $\Lambda \rightarrow 0$). On the other hand, the destruction number $\Gamma_{M}$ is associated with the destruction or breaking down of the fluid microstructure, and quantifies the rate at which the viscosity of the fluid decreases when subjected to flow. A further discussion on the significance of these dimensionless numbers can be found elsewhere~\citep{castillo2018elastic}.

\section{Numerical details}

\subsection{Solution techniques, initial and boundary conditions}

To solve the aforementioned governing equations, namely, mass, momentum, viscoelastic constitutive, and fluidity evaluation equations mentioned in the previous section, the finite volume method-based open-source computational fluid dynamics (CFD) code OpenFOAM~\citep{wellerOpenFOAM} and a recently developed rheoFoam solver available in RheoTool~\citep{rheoTool} have been used in the present study. The time derivative terms were discretised using the first-order Euler scheme, whereas the advective terms in the momentum, constitutive, and fluidity equations were discretised using the high-resolution CUBISTA (Convergent and Universally Bounded Interpolation Scheme for Treatment of Advection) scheme for its improved iterative convergence properties. The diffusion terms in these equations were discretised using the second-order accurate Gauss linear orthogonal interpolation scheme. While the linear systems of the pressure and velocity fields were solved using the Preconditioned Conjugate Gradient (PCG) solver with DIC (Diagonal-based Incomplete Cholesky) preconditioner, the stress fields were solved using the Preconditioned Bi-conjugate Gradient (PBiCG) solver with DILU (Diagonal-based Incomplete Lower-Upper) preconditioner. The pressure-velocity coupling was achieved using the SIMPLE (Semi-Implicit Method for Pressure Linked Equations) method. Furthermore, the relative tolerance levels for the pressure, velocity, and stress fields were set to 10$^{-10}$. The log-conformation tensor approach was used to stabilise the numerical solutions, which was originally proposed by Fattal and Kupferman~\citep{fattal2004constitutive} and first implemented in OpenFOAM by Pimenta and Alves~\citep{pimenta2017stabilization}. In this approach, the fluid microstructure contribution to the stress tensor is first written in terms of conformation tensor $\textbf{A}$ as follows
\begin{equation}
    \bm{\tau}_p = \frac{(1-\beta)}{Re Wi} \left( \textbf{A} - \textbf{I}\right)
\end{equation}
Where the conformation tensor $\textbf{A}$ is evaluated as per the following equation
\begin{equation}
    \frac{\partial \textbf{A}}{\partial t} + \bm{u} \cdot \nabla \textbf{A} - \textbf{A} \cdot (\nabla \textbf{u}) + (\nabla \textbf{u})^{T} \cdot \textbf{A} = - \frac{\phi}{Wi}\left( \textbf{A} - \textbf{I} \right)
    \label{conformation}
\end{equation}
The log-conformation tensor approach involves the introduction of a new tensor $\bm{\Theta}$, which is the natural logarithm of the conformation tensor, as follows
\begin{equation}
    \bm{\Theta}  = \text{ln} \textbf{A}
\end{equation}
Since $\textbf{A}$ is positive definite, it can be diagonalised as $\textbf{A} = \textbf{R} \Upsigma \textbf{R}^{T}$, where $\textbf{R}$ is a matrix that contains in its columns the eigenvectors of $\textbf{A}$ and $\Upsigma$ is a matrix whose diagonal elements are the eigenvalues resulting from the decomposition of $\textbf{A}$. As a result, the new tensor can be written as follows
\begin{equation}
    \bm{\Theta} = \textbf{R} \,\text{In} ( \Upsigma) \textbf{R}^{T}
\end{equation}
Finally, instead of solving the equation~\ref{conformation}, we solve the following equation in terms of this new tensor
\begin{equation}
    \frac{\text{D} \bm{\Theta}}{\text{D}t} = \bm{\Omega} \bm{\Theta} - \bm{\Theta} \bm{\Omega} + 2 \textbf{B} + \frac{\phi}{Wi} \left( \text{exp}(-\bm{\Theta}) - \textbf{I} \right)
\end{equation}
In the above equation, $\bm{\Theta}$ is an anti-symmetric tensor describing the pure rotational component of the velocity gradient tensor and $\textbf{B}$ is a symmetric tensor containing the traceless extensional component of the velocity gradient tensor.

Finally, to complete the problem formulation in the present study, we adopt the following set of boundary and initial conditions 

\textbf{At the inlet patch:} A uniform streamwise velocity of $U_{\infty}$ was utilised. The pressure gradient, along with the extra stress tensor, was set to zero. Moreover, the fluidity field was set to a zero shear-rate value.

\textbf{At the outlet patch:} For all the variables, a Neumann-type boundary condition was adopted except for pressure, for which we provide a zero value.

\textbf{On the surface of cylinder:} The standard no-slip condition for velocity, zero-gradient for pressure and fluidity, and linear extrapolation for extra stress tensor were employed at this solid surface.

\textbf{At the top and bottom symmetry planes:} For all the variables, we use a symmetry type boundary condition at both these fictitious planes.

As an initial condition, the fluidity is specified at a zero shear rate, and the fluid is stress-free throughout the whole computational domain.

\subsection{Grid and time-step convergence studies}

\begin{table}
  \caption{Details of grid and time-step convergence study performed using the MBM model at $Re = 62$. Here, the bold values are those that are finally considered after the grid and time-step convergence studies.}
  \label{table:grid_test}
  \begin{center}
  \begin{tabular}{cccccc}
    \hline
     Grid type & Cells on cylinder surface & Total number of cells & Time-step & $C_D$ & $St$\\[2pt]
    \hline
    G-1 & 300 & 57,105 & $4.312 \times 10^{-4}$ & 1.284 & 0.231 \\
    G-2 & 400 & 102,000 & $4.312 \times 10^{-4}$ & 1.302 & 0.233 \\
    \multirow{2}{*}{\textbf{G-3}} & \multirow{2}{*}{\textbf{500}} & \multirow{2}{*}{\textbf{158,925}} & \bm{$4.312 \times 10^{-4}$} & \textbf{1.318} & \textbf{0.232} \\
     & & & $2.156 \times 10^{-4}$ & 1.321 & 0.233 \\
    G-4 & 600 & 228,420 & $4.312 \times 10^{-4}$ & 1.331 & 0.230 \\[2pt]
    \hline
  \end{tabular}
  \end{center}
\end{table}
To ensure both numerical accuracy and computational efficiency, this study conducts rigorous grid and time-step independence tests. For the grid independence study, four grids, namely G-1, G-2, G-3, and G-4, with progressively refined numbers of cells on the cylinder surface and throughout the computational domain, have been created, as detailed in Table~\ref{table:grid_test}. Simulations were run at the highest $ Re$ value of 62, and results were compared for variations in the time-averaged drag coefficient and Strouhal number at different grid densities. Results hardly change as one moves from G-3 to G-4 (with a difference of less than 1\%), while the total number of cells increases to almost 1.5 times, resulting in a higher computational cost. Therefore, the grid G-3 is chosen for the present study across the full range of conditions. Careful consideration is given when any grid is created. For instance, a structured grid arrangement with hexahedral cells is utilised in this study, wherein cells are progressively concentrated in the vicinity of the cylinder to capture the steep gradients of velocity, stress, and fluidity, as schematically shown in sub-figure~\ref{schematicAndGrid}(b). Additionally, cells are concentrated downstream of the cylinder to capture the flow physics, particularly the vortex dynamics, accurately in this region. To further guarantee temporal accuracy, simulations were conducted using two different time-step sizes on the selected grid. The results were virtually identical, confirming temporal convergence. Consequently, the smaller time step of $\Delta t = 4.312 \times 10^{-4}$, is chosen for all simulations across the full range of Reynolds numbers to ensure robustness and consistency in the present study. 

\subsection{Solver validation}

\begin{figure}
    \centering
    \includegraphics[width=14cm]{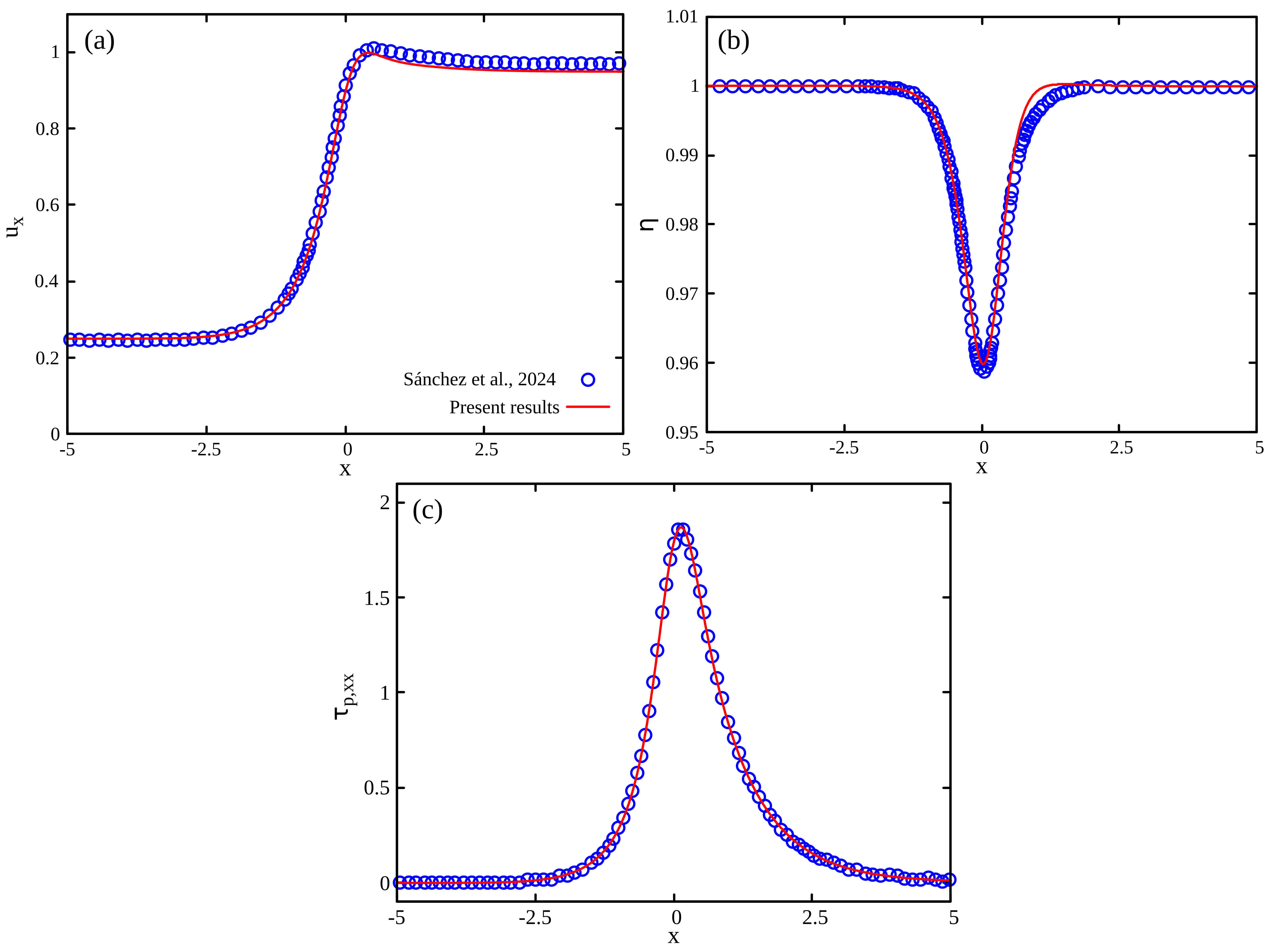}
    \caption{Comparison of the present numerical results with the corresponding prior results of Castillo-S{\'a}nchez et al.~\citep{castillo2024numerical} for flow through a planar 4:1 contraction geometry. The parameters used here are for a fixed fluid with strong hardening behaviour (structural relaxation parameter, $\Lambda = 0.28$), destruction parameter, $\Gamma_M = 0.1125$, Weissenberg number, $Wi = 1$, and Reynolds number, $Re = 1.11$. Here, sub-figures(a), (b), and (c) represent the comparison of streamwise velocity, apparent viscosity, and streamwise stress component along a horizontal line passing through the middle of the geometry.}
    \label{validation1}
\end{figure}

\begin{figure}
    \centering
    \includegraphics[width=14cm]{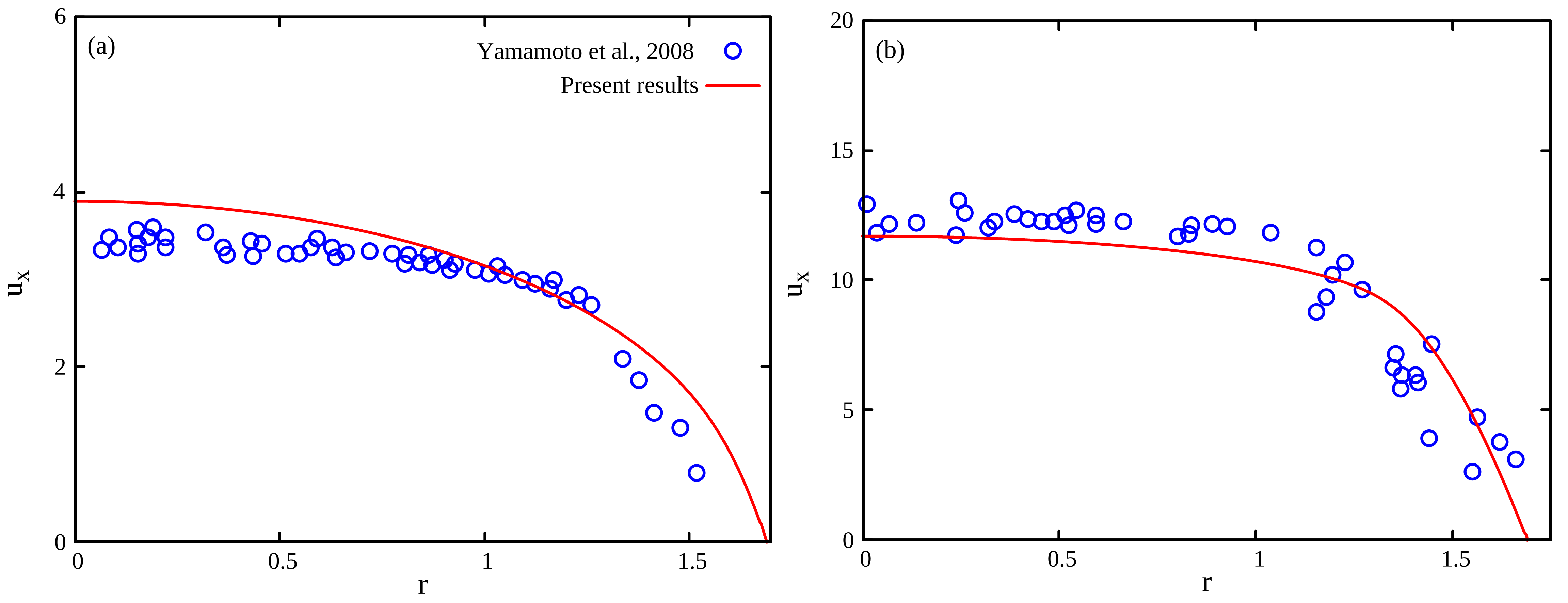}
    \caption{Comparison of axial velocity profile of wormlike micellar solutions flowing through an axisymmetric capillary between the present numerical simulations based on the MBM model and the corresponding experimental results of Yamamoto et al.~\citep{yamamoto2008flow} at two different inlet velocities, namely, (a) $u_0 = 2.55 \,mm/s$ and (b) $u_0 = 8.67\, mm/s$. Here, the MBM model parameters used to carry out the present simulations are~\citep{hashimoto2006effects}: $G_0 = 5.1\,Pa$, $\lambda = 0.6 \,s$, $\eta_{s} = 0.1 \,Pa \cdot s$, $\frac{1}{\phi_{\infty}} = \eta_{\infty} = 0.08 \,Pa \cdot s$, $\frac{1}{\phi_{0}} = {\eta_{p}}_0 = 3.1 \, Pa \cdot s$, and $k_0 = 0.001\,Pa^{-1}$.}
    \label{validation2}
\end{figure}

To establish the accuracy and reliability of the present numerical solver, validation studies have been conducted using prior numerical and experimental results. Figure~\ref{validation1} shows the comparison between the present numerical results with those of Castillo-S{\'a}nchez et al.~\citep{castillo2024numerical} on flows through an expansion-contraction microchannel (4:1 ratio) wherein the results are obtained with the MBM model. In particular, results are compared with respect to the spatial variation of various quantities, such as the streamwise velocity component, apparent viscosity, and stress component, along a horizontal line passing through the middle of the microchannel. A very good agreement is observed between the two results in figure~\ref{validation1}. Yamamoto et al.~\citep{yamamoto2008flow} performed an experimental study on the flow dynamics of wormlike micellar solutions, consisting of cetyltrimethylammonium bromide (CTAB) with sodium salicylate (NaSal), through an axisymmetric capillary and measured the velocity profiles using the particle tracking velocimetry technique at various flow rates. Figure~\ref{validation2} displays the comparison of the axial velocity profile between the present numerical simulations using the MBM model and the corresponding experimental results of Yamamoto et al. at two different inlet velocities. Note that the MBM model parameters used for this validation study have already been provided by them, which they obtained by fitting the steady shear viscosity data and elastic and loss moduli data obtained from small amplitude oscillatory shear experiments~\citep{hashimoto2006effects}. A fairly good agreement is observed between the numerical and experimental results in figure~\ref{validation2}. In particular, the present MBM model successfully captures the plateau region of the velocity profile in the middle of the capillary and the sudden steep gradient near the capillary wall.      

\section{Results and discussion}
The simulations are carried out for the MBM model for structured fluids and the simple Newtonian fluid model to compare the flow physics under otherwise identical conditions. The MBM model parameters used in this study are obtained by fitting the experimental results of steady shear and small amplitude oscillatory shear test results provided by Raghavan and Kaler~\citep{raghavan2001highly} for erucyl bis(hydroxyethyl)methylammonium chloride (EHAC) and sodium salicylate (NaSal) micellar system with a molar ratio of 0.65 at 25$^\circ$C, as demonstrated in Appendix~\ref{App:A}. The model parameters are: $\lambda = 0.323 \,s$, $G_0 = 50.02 \,Pa$, $\frac{K_0}{\eta_\infty} = 0.323$, $\eta_{s} = 0.001 \,Pa \cdot s$, ${\eta_{p}}_0 = 6.955\,Pa \cdot s$, and $\rho = 1000 \, kg/m^{3}$. After substituting these values in the dimensionless numbers, we get the following ranges of values of those numbers, namely, $Re = 30 -62$, $Wi = 0.029-0.06$, $\Lambda = 0.067-0.139$, $\Gamma_{M} = 3.260-6.738$, $\beta = 0.000144$. Note that here the dimensionless numbers are varied by changing $U_{\infty}$ imposed at the inlet patch of the computational domain, and the rest of the parameters remain fixed. Consequently, only the Reynolds number variation is reported, which is of particular interest in the current study. The results are presented and discussed in terms of various flow features, including streamlines and velocity magnitude, vorticity patterns, variations in surface-integral parameters, dynamic mode decomposition analysis, etc.

\subsection{Transition in flow field with increasing Reynolds number}

\begin{figure}
    \centering
    \includegraphics[width=14cm]{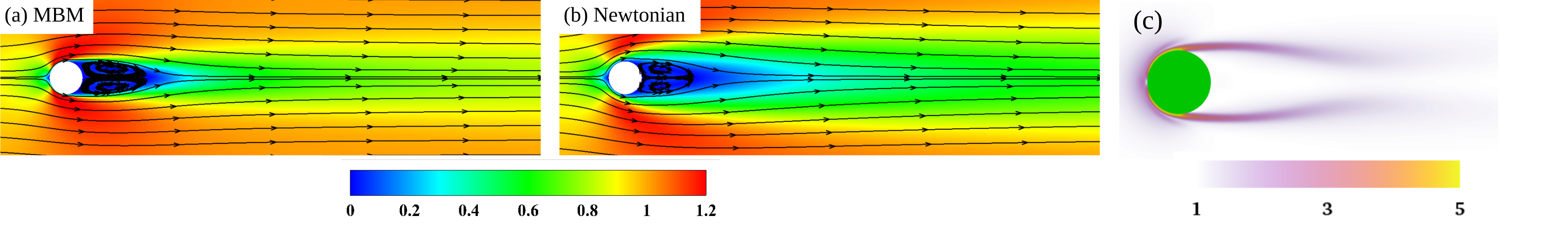}
    \caption{Streamlines and velocity magnitude plots at $Re = 30$ for micellar solutions (a) and Newtonian fluids (b) at the same Reynolds number of 30. The distribution of fluidity $\phi$ for micellar solutions is shown in sub-figure (c).}
    \label{Stream_Re30}
\end{figure}

Although this study primarily aims to investigate the unsteady flow regime, some simulations are also conducted in the steady flow regime, in which recirculation zones form but remain attached to the cylinder surface. Figure~\ref{Stream_Re30} shows the streamlines and velocity magnitude patterns at $Re = 30$, both for wormlike micellar solutions and simple Newtonian fluids. At this flow condition, two equal-sized clockwise and counter-clockwise rotating recirculation zones are formed downstream of the cylinder in Newtonian fluids (sub-figure~\ref{Stream_Re30}(b)), as it was established in several experimental and numerical studies~\citep{zdravkovich1997flow,williamson1996k}. The same trend is also seen for structured fluids. However, there are noticeable differences between them. For instance, the low-velocity magnitude zone is more concentrated downstream of the cylinder in micellar solutions than in Newtonian fluids, sub-figure~\ref{Stream_Re30}(a). Furthermore, the recirculation zones are longer in micellar solutions than in Newtonian fluids. The separation angle (measured from the cylinder's front stagnation point) of the boundary layer forming the downstream wake is smaller in the former solution than in the latter. This suggests that an early boundary-layer separation occurs in micellar solutions than in Newtonian fluids. This is because micelles break in the vicinity of the cylinder due to the high shear in this region, thereby lowering the apparent viscosity $\eta_{\text{app}}$ (i.e., increasing fluidity), as evident in sub-figure~\ref{Stream_Re30}(c). This reduction in apparent viscosity weakens viscous momentum diffusion in the y-direction from outer faster fluid layer to inner slower fluid layer $\left( \sim \frac{\partial}{\partial y} \left(\eta_{\text{app}}  \frac{ \partial u_y}{\partial y}\right) \right)$ and wall shear stress, as both of which are proportional to the local viscosity, making the boundary layer unable to overcome the adverse pressure gradient, and hence, it separates earlier.  Furthermore, it can be seen from sub-figure~\ref{Stream_Re30}(c) that a strand of low viscosity zone is formed along the edge of the reciculation zone (due to breakage of micelles because of the high extensional flow field in this region), which again inhibits the momentum transfer from the outside layer of the reciculation zone to its inside. This is why the low-velocity magnitude zone is concentrated in the wake region of micellar solutions. Furthermore, it delays velocity recovery in the wake, thereby increasing its downstream length in micellar solutions. 

\begin{figure}
    \centering
    \includegraphics[width=14cm]{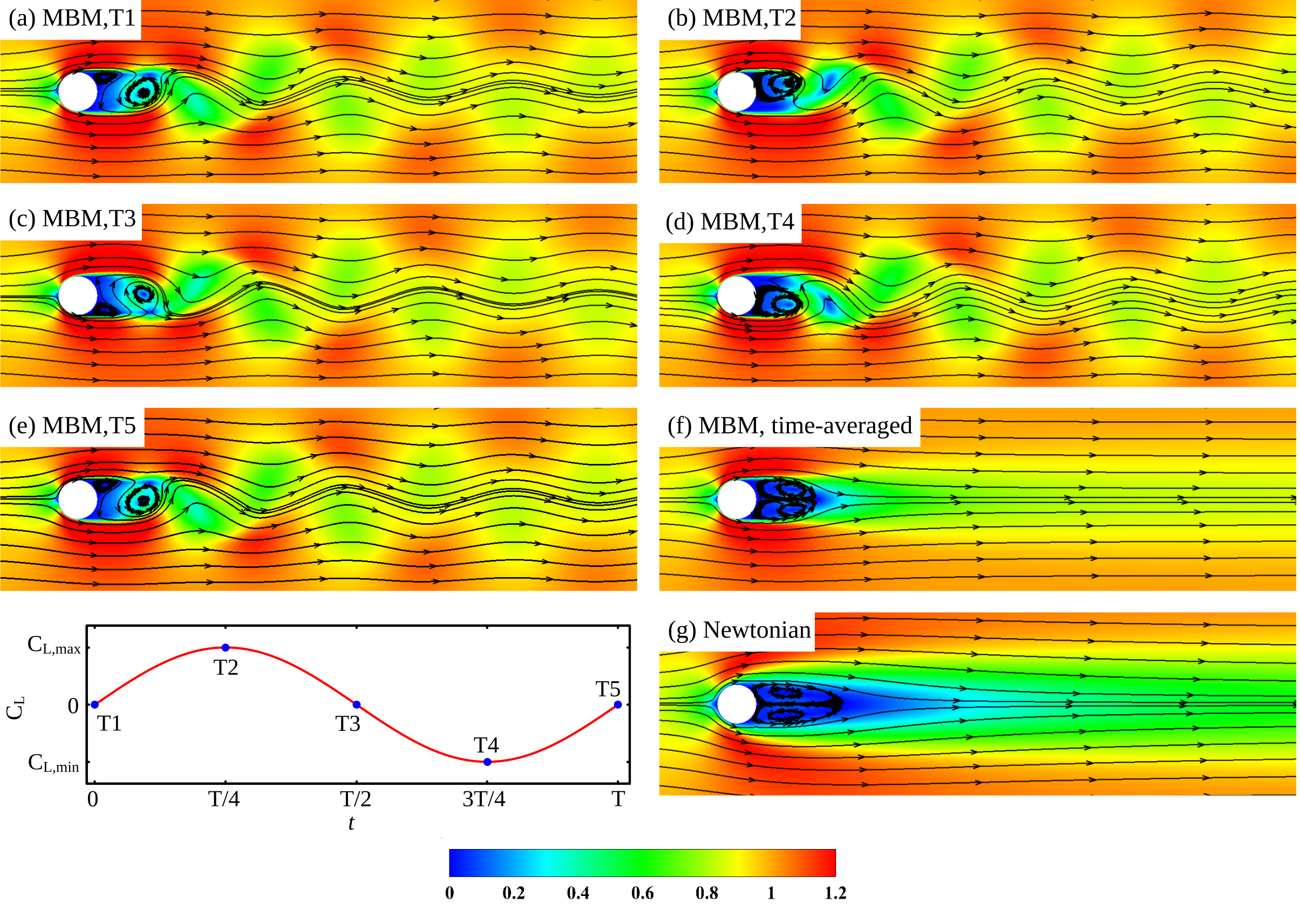}
    \caption{Streamline patterns and velocity magnitude plots at $Re = 40$ both for micellar solutions (a-e) and Newtonian fluids (g). The time-averaged streamlines and velocity magnitude plots are shown in sub-figure (f). The bottom-left sub-figure shows different time instances at which the flow fields are shown for micellar solutions.}
    \label{Fig:Stream_Re40}
\end{figure}

As the Reynolds number increases to 40, the wake length increases in Newtonian fluids, but it remains attached to the cylinder, as shown in sub-figure~\ref{Fig:Stream_Re40}(g), due to an increase in inertial forces. This is again a well-known finding both in experiments and simulations~\citep{williamson1996k,zdravkovich1997flow}. However, strikingly, a different flow phenomenon occurs in micellar solutions under the same flow conditions. In these structured fluids, the wake is no longer attached to the cylinder; instead, it starts to shed from the cylinder surface, for instance, see the results in sub-figures~\ref{Fig:Stream_Re40}(a-e) for five different time instants (T1--T5), as shown in the bottom left corner of figure~\ref{Fig:Stream_Re40}. At time instant T1, a wall-attached clockwise recirculation zone forms at the upper downstream corner of the cylinder, whereas an anti-clockwise one forms at a distance almost twice the cylinder radius from its rear stagnation point. In the next time instant T2, the wall-attached recirculation zone grows in size, whereas the other one formed away from the cylinder shrinks. The upper corner recirculation zone is ultimately detached from the cylinder surface, and an anti-clockwise recirculation zone appears at the bottom corner of the cylinder at time instant T3. The latter one grows in size and detaches from the cylinder surface at T4. A recirculation zone forms at the upper corner of the cylinder at T5 and repeats the flow structure as seen at T1. The corresponding time-averaged flow structure is shown in sub-figure~\ref{Fig:Stream_Re40}(f), and it can be seen that the time-averaged recirculation zone is smaller than that observed for Newtonian fluids. Furthermore, a clear separation between the high- and low-velocity magnitude zones around the cylinder and in the downstream wake is evident at this Reynolds number, as in $Re = 30$ (figure~\ref{Stream_Re30}). 

\begin{figure}
    \centering
    \includegraphics[width=14cm]{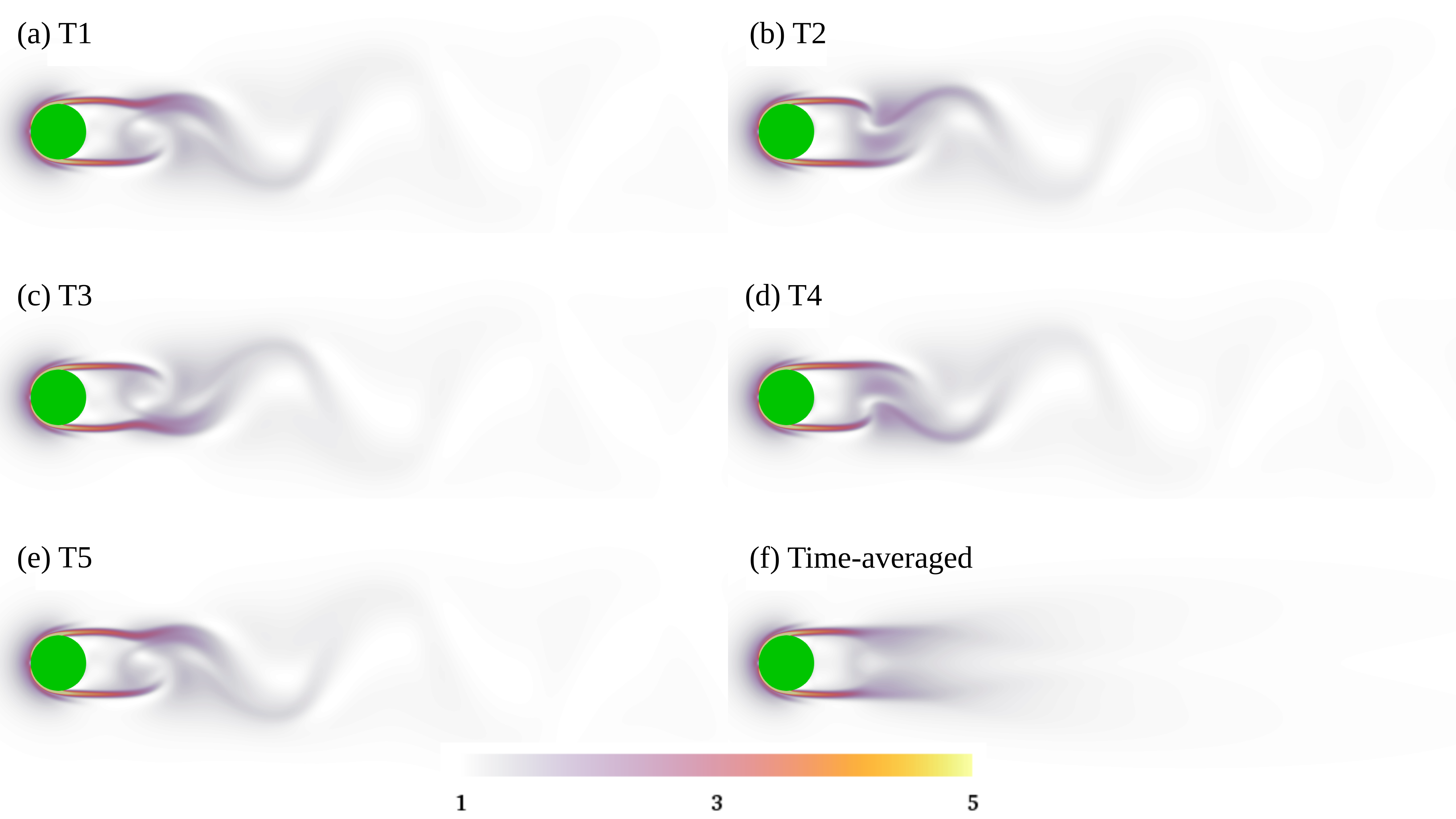}
    \caption{Fluidity patterns at $Re = 40$ for micellar solutions at different time instants. The corresponding time-averaged patterns are shown in sub-figure (f).}
    \label{Fluidity_Re40}
\end{figure}

This time variation and repeating pattern of the flow field suggest that a transition from steady to unsteady flow occurs in micellar solutions at $Re = 40$, whereas it remains steady in Newtonian fluids under the same flow conditions. This early transition in the flow field could be attributed to the Kelvin-Helmholtz instability, a shear-layer instability that occurs at the edges of the recirculation zone. The likelihood of these instabilities is higher in micellar solutions, as the velocity gradient normal to the flow is very steep at the edges of the recirculation zone, as shown in the velocity magnitude distribution downstream of the cylinder. As mentioned earlier, this stark distribution in velocity magnitude is caused by micelle breakage, which increases fluidity and reduces apparent viscosity. This, in turn, inhibits momentum diffusion in the normal direction to the flow. The corresponding surface plot of fluidity at this Reynolds number and different time instants further confirms the presence of high-fluidity strands at the wake edges downstream of the cylinder (figure~\ref{Fluidity_Re40}). These strands become more pronounced at $Re = 40$ compared to those seen at $Re = 30$, due to the breakage of more micelles caused by increased inertial forces. This ultimately triggers instability and leads to the early onset of vortex shedding. A detailed discussion on this is, once again, presented in subsection~\ref{mechanism_flow_transition}. 

\begin{figure}
    \centering
    \includegraphics[width=14cm]{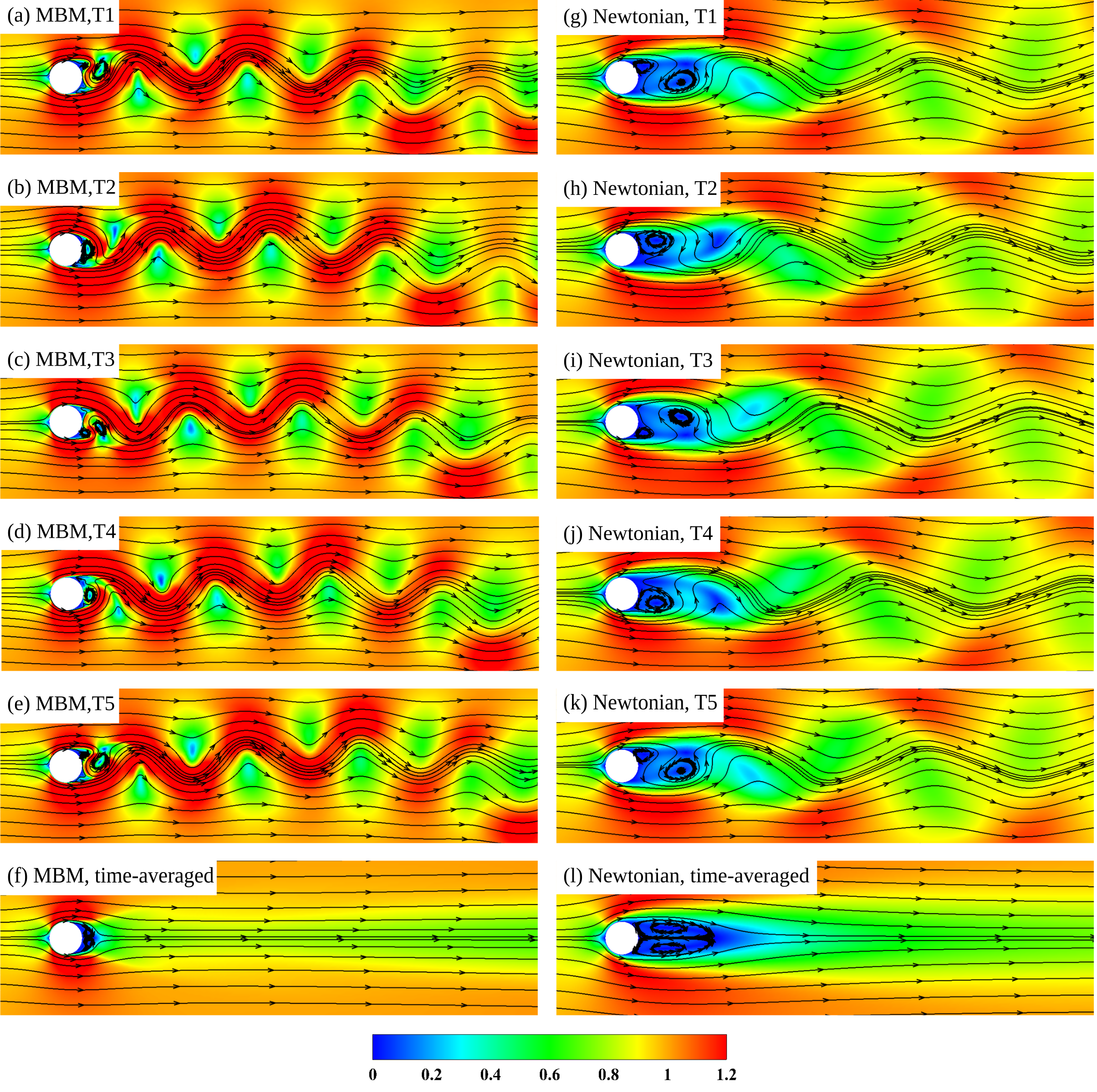}
    \caption{Streamline patterns and velocity magnitude plots at $Re = 62$ both for micellar solutions (a-e) and Newtonian fluids (g-k). The corresponding time-averaged streamlines and velocity magnitude plots are shown in sub-figure (f) and sub-figure (l), respectively.}
    \label{Stream_Re62}
\end{figure}

At $Re = 62$, Newtonian fluids exhibit a similar flow pattern as that seen for micellar solutions at $Re = 40$, i.e., the formation of clockwise and anti-clockwise recirculation zones at the cylinder's upper and lower rear sides and their subsequent increase in size and then ultimately detachment from the cylinder surface at different time instants of the cycle, figure~\ref{Stream_Re62}. Micellar solutions also display a similar trend. However, the corresponding recirculation zones are much smaller than those observed in Newtonian fluids and in the same solution at $Re = 40$. Furthermore, multiple recirculation zones are now observed in the vicinity of the cylinder surface simultaneously in micellar solutions. For instance, at time instant T1 (sub-figure~\ref{Stream_Re62}(a)), two clockwise-rotating recirculation zones are observed at the cylinder's upper and lower rear sides simultaneously, and an anti-clockwise recirculation zone away from the cylinder is seen at the same time. At the next time instant T2, these wall-bounded recirculation zones merge to form a larger recirculation zone, which ultimately detaches from the cylinder surface at T3. The time-averaged plot is shown in sub-figure~\ref{Stream_Re62}(f), where several smaller recirculation zones are seen to be present compared to two large recirculation zones seen for Newtonian fluids in sub-figure~\ref{Stream_Re62}(l). Furthermore, due to the smaller recirculation zones formed downstream of the cylinder, the low-velocity magnitude zone there decreases drastically in micellar solutions compared to Newtonian fluids.       

\begin{figure}
    \centering
    \includegraphics[width=14cm]{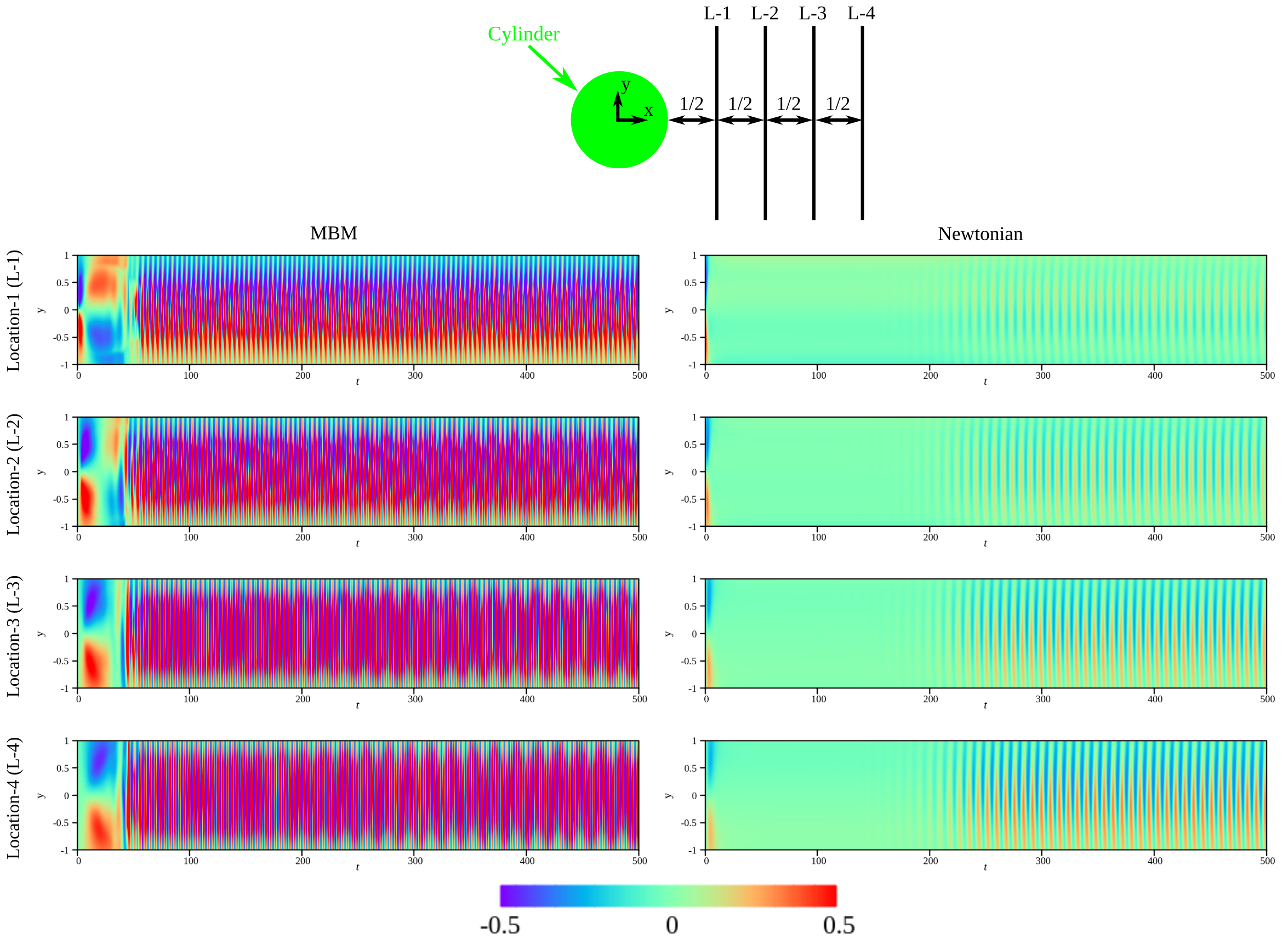}
    \caption{Kymograph for spanwise velocity component at four different vertical lines, namely, location-1 (L-1): $x = 1$, location-2 (L-2): $x = 1.5$, location-3 (L-3): $x = 2$, and location-4 (L-4): $x = 2.5$. The results are shown here for both micellar solutions (left half) and Newtonian fluids (right half) at $Re = 62$.}
    \label{kymograph}
\end{figure}

\begin{figure}
    \centering
    \includegraphics[width=13cm]{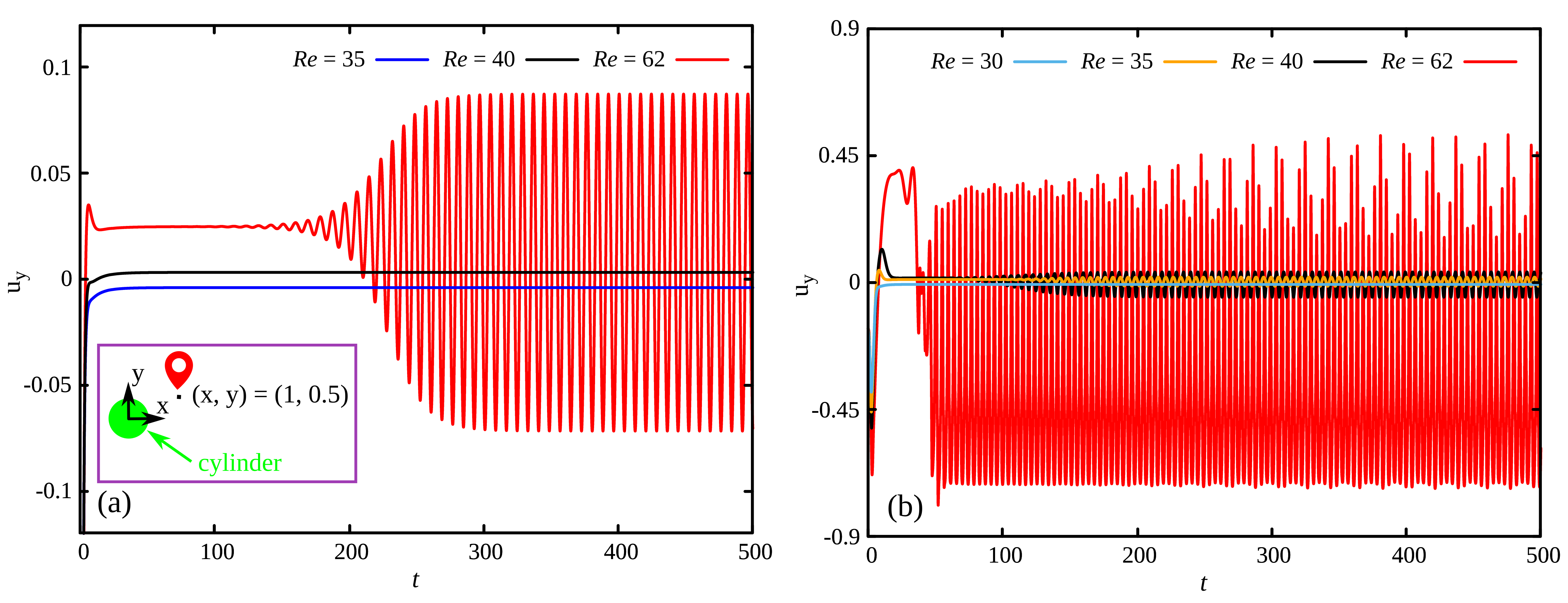}
    \caption{The temporal variation of the spanwise velocity component at a probe location (x = 1, y = 0.5) at different Reynolds numbers for Newtonian (a) and micellar solutions (b).}
    \label{Probe}
\end{figure}

\begin{figure}
    \centering
    \includegraphics[width=13cm]{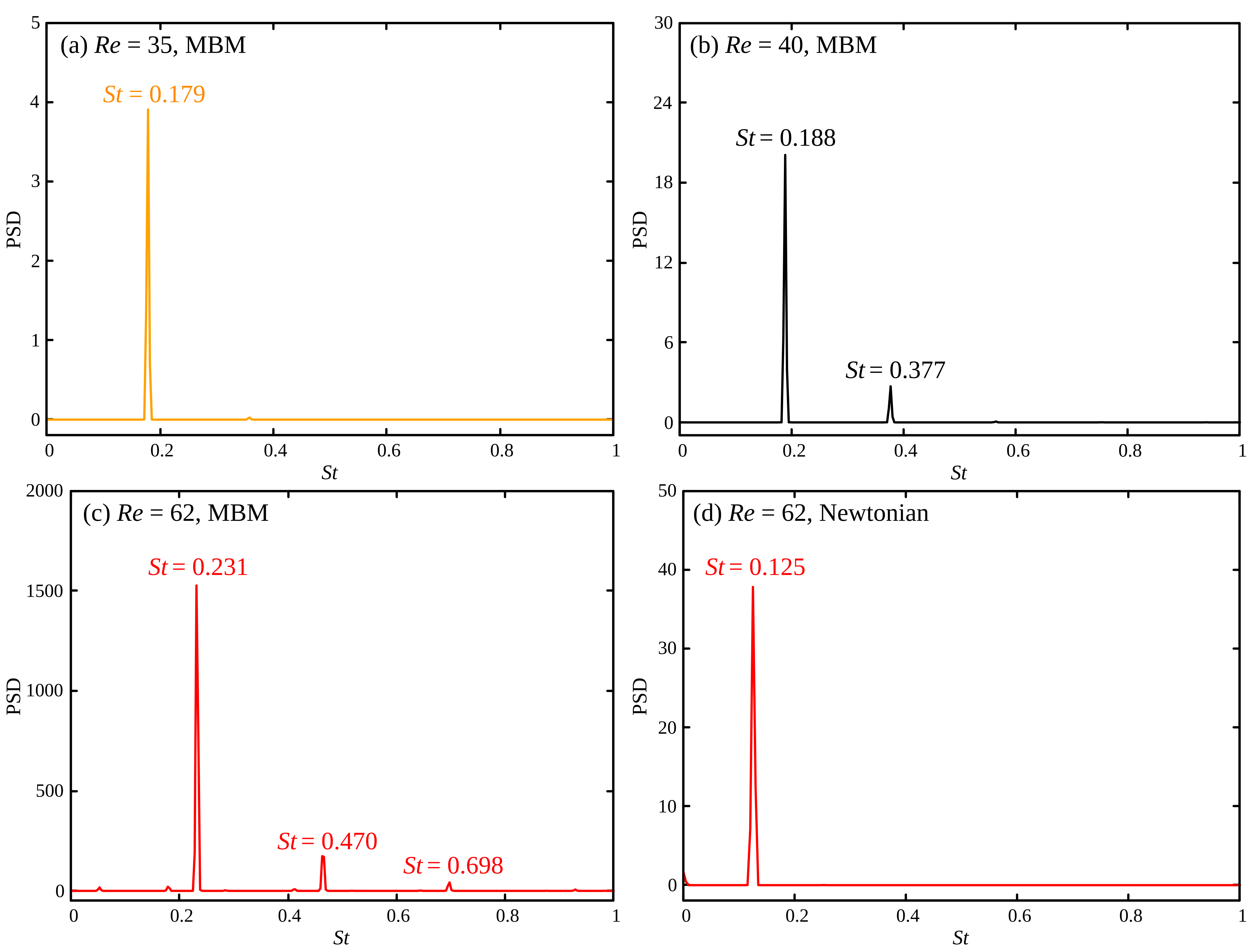}
    \caption{The power spectral density plot of spanwise velocity fluctuations at different Reynolds numbers both for micellar solutions and Newtonian fluids. Here, only the unsteady cases observed in figure~\ref{Probe} are shown.}
    \label{PSD}
\end{figure}

It is now confirmed that the flow field in both fluids becomes unsteady at this Reynolds number. To know the nature of this unsteady flow field, a kymograph of the spanwise velocity component is plotted at four different vertical locations downstream of the cylinder at $Re = 62$, figure~\ref{kymograph}. In Newtonian fluids, the transition from steady to unsteady flow takes time, and once it occurs, the flow field exhibits a regular, periodic structure with a dominant and steep spike. In contrast, the transition is much faster in micellar solutions, and the flow field also exhibits a spectrum of spikes over time, suggesting a quasi-periodic flow field. Furthermore, the fluctuations are more intense in micellar solutions than in Newtonian fluids. To gain more insights into the flow field, the temporal variation of the spanwise velocity component $u_y$ at a probe location downstream of the cylinder is depicted in figure~\ref{Probe} for both Newtonian and micellar solutions. For Newtonian fluids, the velocity reaches a steady value with time at $Re = 35$ and 40, whereas for micellar solutions, it reaches a steady value at $Re = 30$ but shows temporal fluctuations at $Re = 35$, 40 and 62. This again confirms that the flow field transitions to an early unsteady state in micellar solutions, compared to that in Newtonian fluids. To elucidate further insights into the flow behaviour, the power spectral density (PSD) plot of velocity fluctuations is presented in figure~\ref{PSD}. From this figure, it can be seen that a single dominant peak in the PSD value is present at around non-dimensional frequency and/or Strouhal number $St (= \frac{f D_{0}}{U_{\infty}}) \sim 0.179$ at $Re = 35$ for micellar solutions (sub-figure~\ref{PSD}(a)), suggesting the presence of an unsteady regular periodic flow behaviour at this condition. As the Reynolds number increases to 40, once again, a dominant peak in the PSD value is present at $St_1 \sim 0.188$ along with a small harmonic peak at $St_{2} \sim 0.377$, wherein $St_{2} \sim 2St_{1}$ (sub-figure~\ref{PSD}(b)). This indicates that the flow remains periodic; however, some nonlinear phenomena are emerging in the flow field, possibly due to micelle breakage and reformation. As the Reynolds number further increases to 62, a dominant peak with a very high PSD value appears, along with several harmonic and sub-harmonic peaks, as can be seen from sub-figure~\ref{PSD}(c). This suggests that micellar solutions transit to a quasi-periodic state at this Reynolds number. In contrast, the flow field remains in the regular periodic state for Newtonian fluids at the same Reynolds number, as can be evidenced from a single dominant peak in the PSD plot at $St \sim 0.125$ in sub-figure~\ref{PSD}(d).                  

\begin{figure}
    \centering
    \includegraphics[width=12cm]{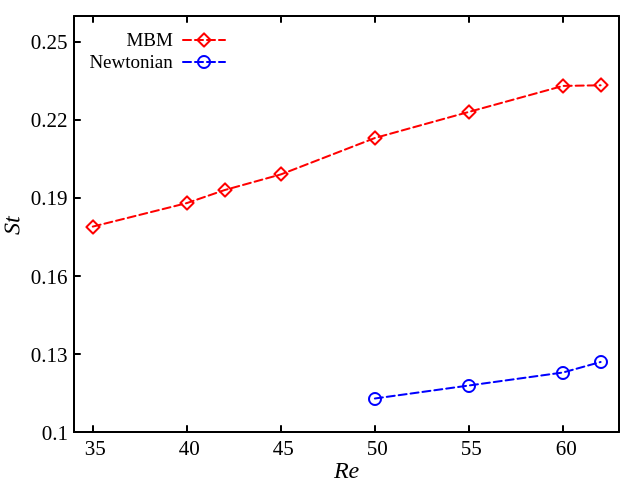}
    \caption{Variation of the Strouhal number with the Reynolds number, both for micellar solutions and Newtonian fluids.}
    \label{StVsRe}
\end{figure}

Figure~\ref{StVsRe} shows the variation of the Strouhal number corresponding to the largest PSD amplitude against the Reynolds number, both for Newtonian fluids and micellar solutions. It can be seen that $St$ values increase with Reynolds number both for Newtonian fluids and micellar solutions. However, the values are much larger, almost double, for micellar solutions than those seen for Newtonian fluids under the same flow conditions. This suggests that the flow phenomena, particularly the shedding of wakes from the cylinder downstream side, are much faster in micellar solutions than in Newtonian fluids.

\subsection{Vorticity patterns and transport rate}

\begin{figure}
    \centering
    \includegraphics[width=14cm]{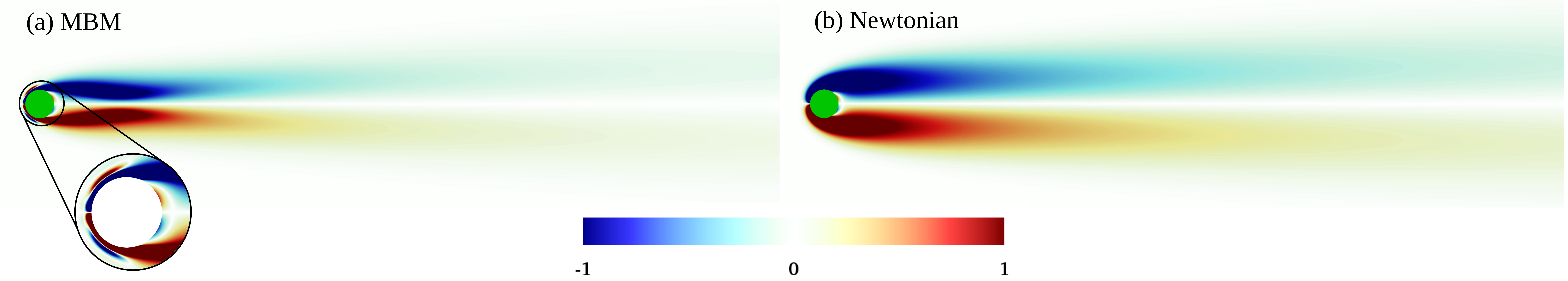}
    \caption{Vorticity patterns for micellar solutions (a) and Newtonian fluids (b) at $Re = 30$.}
    \label{vorticity_Re30}
\end{figure}

The $z$-direction vorticity field $(\omega)$ is presented in figure~\ref{vorticity_Re30} both for Newtonian fluids and micellar solutions at $Re = 30$. The positive vorticity or anti-clockwise rotating vorticity is shown in red colour, whereas the negative vorticity or clockwise rotating velocity is presented in blue colour. The vorticity is generated in the upstream region of the cylinder, travels along its surface, and is ultimately convected into the shear layer downstream of the cylinder. At the rear of the cylinder, two small base region vortices form and remain attached to the cylinder. The upper rear side one is called the negative base region vorticity (NBRV), whereas the lower one is called the positive base region vorticity (PBRV). In micellar solutions, the shear layer vortices are more squeezed at the upstream of the cylinder and less thick downstream of the cylinder than in Newtonian fluids. Furthermore, two small vortex regions appear outside the shear-layer vortex regions at the cylinder upstream in micellar solutions (see the zoomed-in figure), but not in Newtonian fluids.

\begin{figure}
    \centering
    \includegraphics[width=14cm]{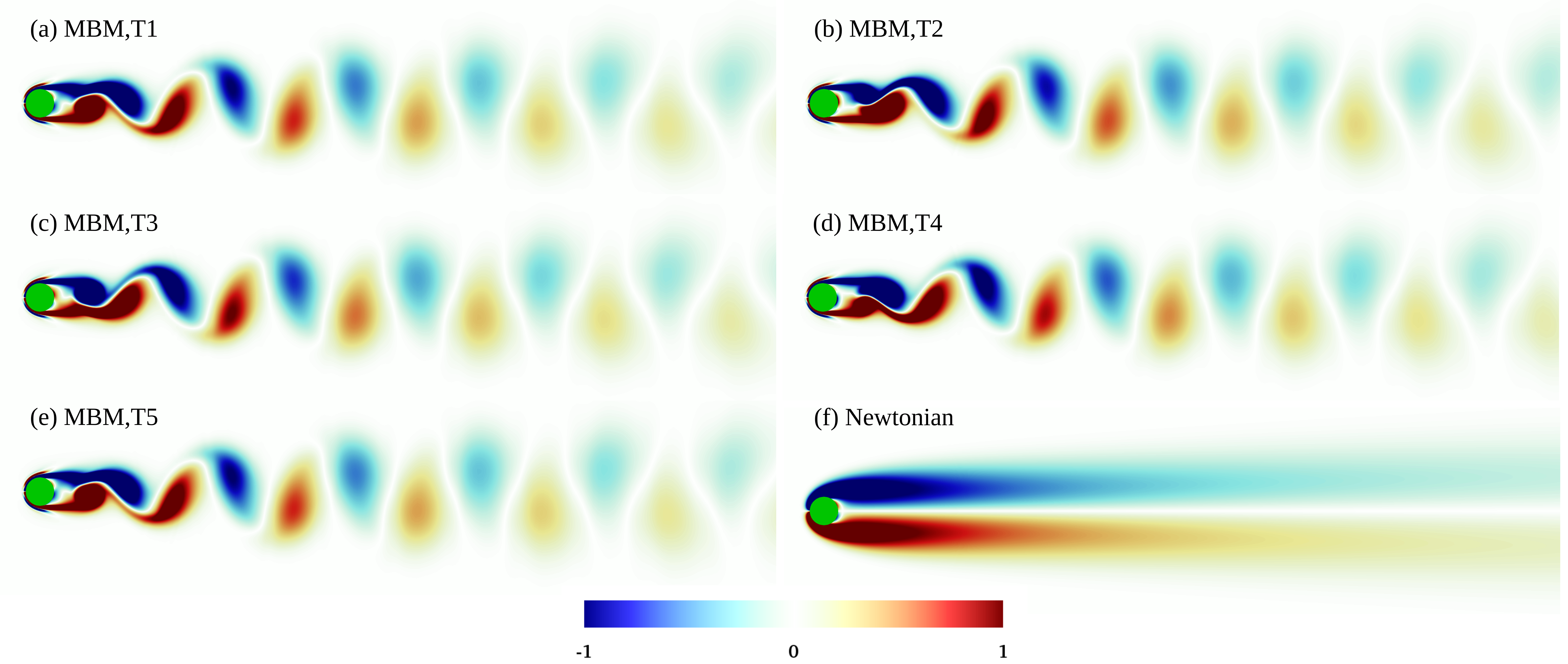}
    \caption{Vorticity patterns for micellar solutions (a-e) and Newtonian fluids (f) at $Re = 40$.}
    \label{vorticity_Re40}
\end{figure}

As the Reynolds number increases to 40, the same pattern is seen for Newtonian fluids as that seen at $Re = 30$. However, a completely different vortex pattern is observed in micellar solutions. In particular, in the latter solutions, the vortices are shed from the cylinder surface, whereas they remain attached in Newtonian fluids, as shown in figure~\ref{vorticity_Re40}. This again confirms the early transition in the flow field from a steady to an unsteady one in micellar solutions. The vortex shedding mechanism in micellar solutions proceeds as follows: at time instant T1 (sub-figure~\ref{vorticity_Re40}(a)), a small NBRV forms at the cylinder's upper rear side, which grows in size at the next time instant T2 (sub-figure~\ref{vorticity_Re40}(b)). It then merges with the lower negative shear-layer vortex and detaches from the cylinder surface at T3, as seen in sub-figure~\ref{vorticity_Re40}(c). At the same time, a small PBRV appears at the cylinder's lower rear side, which then grows in size in the next time instant T4 (sub-figure~\ref{vorticity_Re40}(d)). This again merges with the upper positive shear layer vortex and detaches from the cylinder surface at T5 (sub-figure~\ref{vorticity_Re40}(e)). At the same time, a small NBRV again appears on the cylinder's upper rear side. After merging with the base-region vortices, the shear-layer vortices grow in size and are broken into smaller vortices due to shear-layer instability, which are shed downstream of the cylinder. Therefore, the formation of base-region vortices at the cylinder rear side, their merging with shear-layer vortices and detachment from the cylinder surface, the growing and breaking of shear-layer vortices due to shear-layer instabilities, and the ultimate detachment from the shear-layer constitute the overall mechanism of vortex shedding. The vortices shed downstream of the cylinder form the well-known von Kármán street, whose width increases with the downstream distance. The centres of these counter-rotating vortices in this street are not in a straight line; instead, they are in a zig-zag pattern. Furthermore, throughout the process, two small vortex regions are observed outside the shear-layer vortices at the cylinder upstream, which are now larger than those seen at $Re = 30$. Once again, such vortices are not present in Newtonian fluids.           

\begin{figure}
    \centering
    \includegraphics[width=14cm]{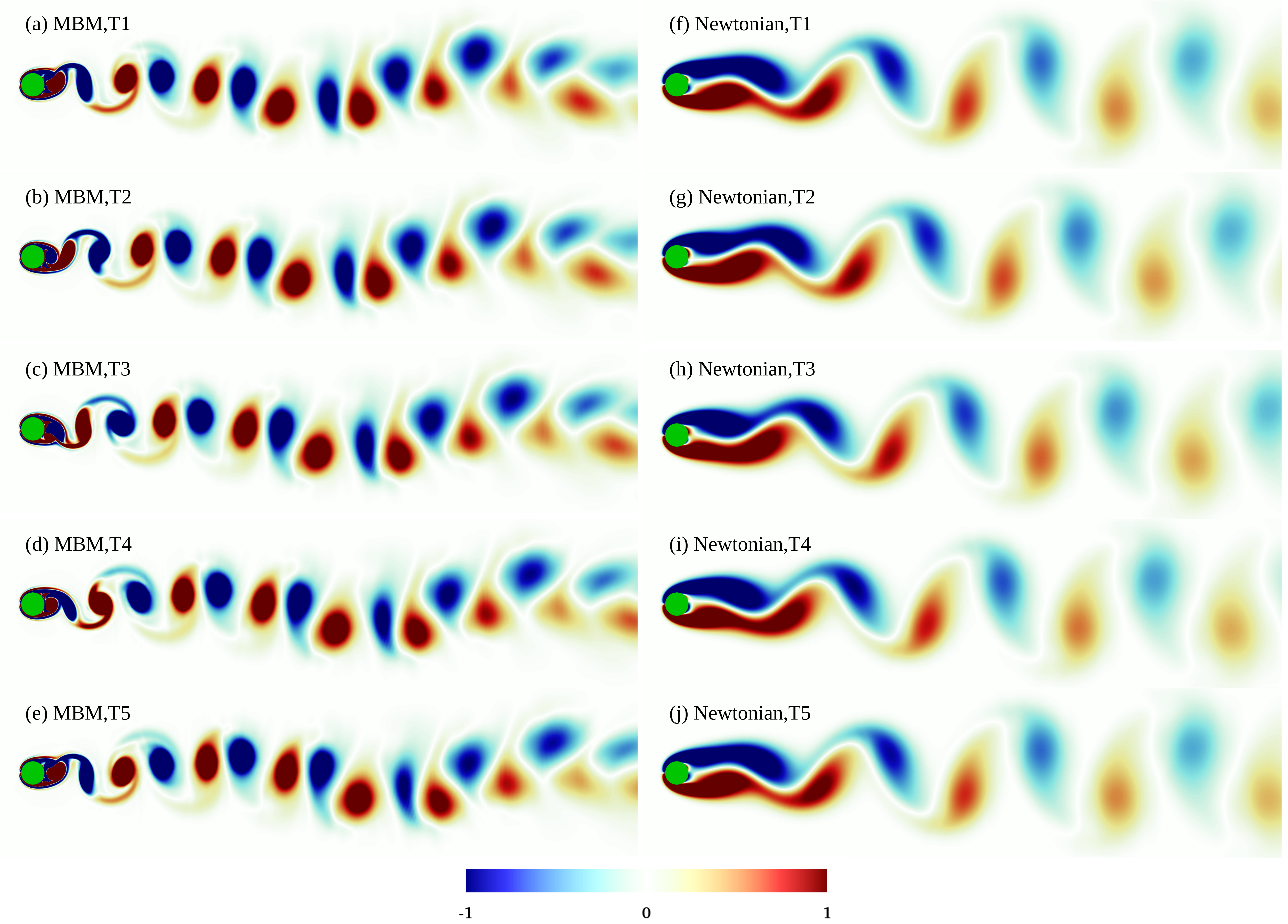}
    \caption{Vorticity patterns for micellar solutions (a-e) and Newtonian fluids (f-j) at $Re = 62$.}
    \label{vorticity_Re62}
\end{figure}

With the Reynolds number further increased to 62, vortex shedding is also observed in Newtonian fluids. This is expected, as the transition in the flow field from steady to unsteady periodic occurs at $Re \sim 49$ for Newtonian fluids~\citep{zdravkovich1997flow,williamson1996k}, which is much lower than $Re = 62$. Although both fluids exhibit vortex shedding at this Reynolds number, the patterns differ substantially. For instance, the vortex street becomes curvy in micellar solutions, whereas it is straight in Newtonian fluids. The downstream vortices are more intense and less diffuse in micellar solutions, extending farther downstream than in Newtonian fluids, resulting in a larger vortex street. Furthermore, they are more deformed and irregular in shape in micellar solutions than in Newtonian fluids. The distance between successive vortices is greater in Newtonian fluids than in micellar solutions. The shear-layer vortices are much thicker and larger in Newtonian fluids than in micellar solutions. The vortices are more concentrated and occupy almost the whole space in the vicinity of the cylinder in micellar solutions. Furthermore, the shear-layer vortices are now wrapped by another layer of vortices in micellar solutions, which were observed to form upstream of the cylinder at lower Reynolds numbers. Such deformation, irregularity, and concentration of vortices downstream of the cylinder in micellar solutions may suggest another flow transition from regular periodic to quasi-periodic flow in the field. Also see the supplementary video provided on this vorticity pattern for better visualisation of the vortex shedding process. 

\begin{figure}
    \centering
    \includegraphics[width=14cm]{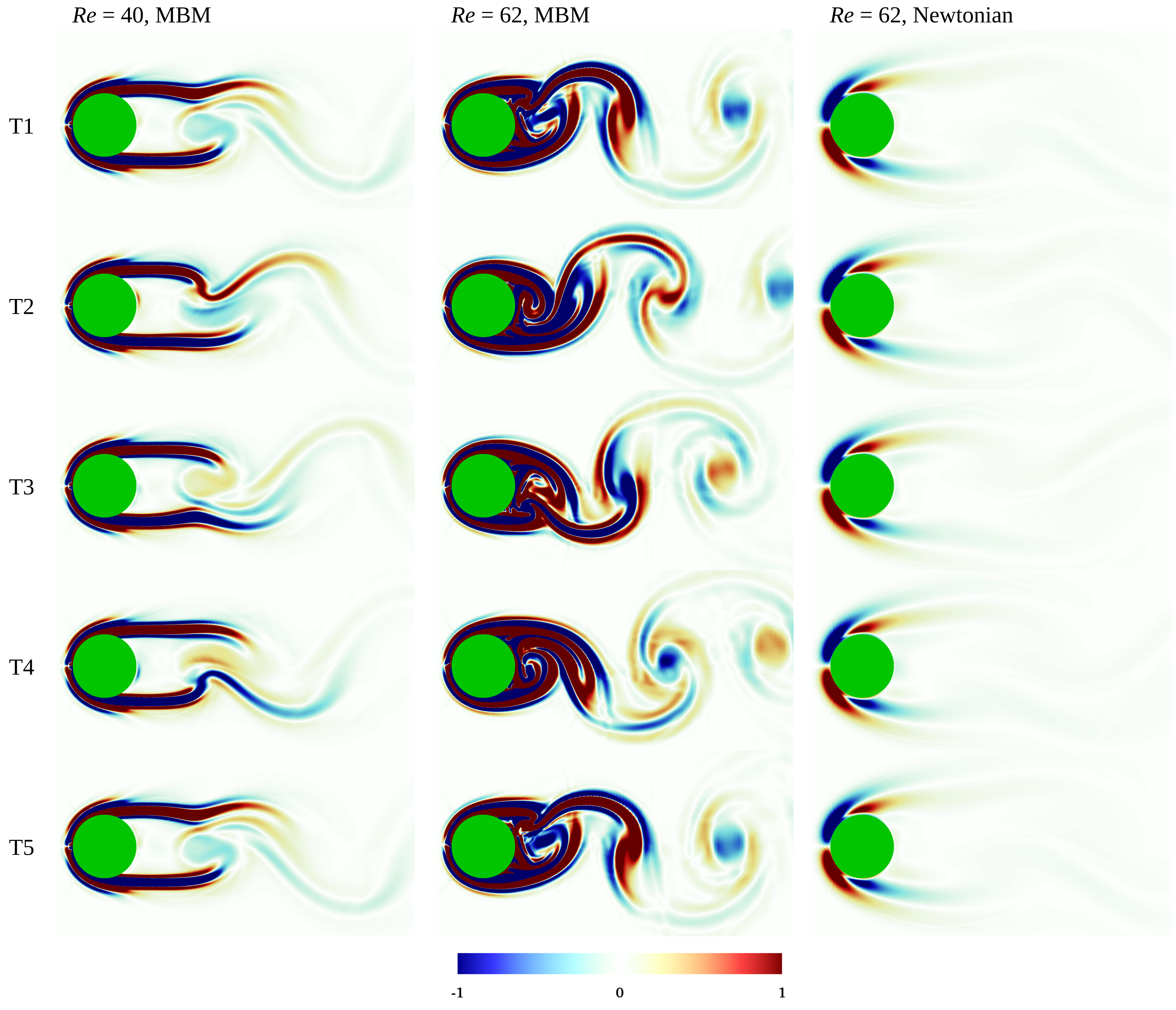}
    \caption{Vorticity transport rate in micellar solutions at $Re = 40$ (first column) and 62 (second column), and in Newtonian fluids at $Re=62$ (third column).}
    \label{vorticityTansport_Re62}
\end{figure}

The vorticity transport rate (VTR) is calculated in the present study as follows

\begin{equation}
    \frac{D \omega}{Dt} = \bm{\nabla} \cdot(\nu \bm{\nabla} \omega) + \frac{\left[\bm{\nabla} \times (\bm{\nabla} \cdot \bm{\tau_{p}}) \right]_{z}}{\rho}
\end{equation}
Where $\omega$ is the vorticity defined as $\omega = \frac{\partial u_{y}}{\partial x} -\frac{\partial u_{x}}{\partial y} $ and $\nu$ is the kinematic viscosity defined as $\nu = \frac{1}{\rho \phi}$. After substituting it and differentiating using the chain rule, we get

\begin{equation}
     \frac{D \omega}{Dt} = \frac{1}{\rho \phi} \bm{\nabla}^{2} \omega - \frac{1}{\rho \phi^{2}} \left(\bm{\nabla} \phi \cdot \bm{\nabla} \omega \right) + \frac{\left[\bm{\nabla} \times (\bm{\nabla} \cdot \bm{\tau_{p}}) \right]_{z}}{\rho}
\end{equation}

The first term on the right-hand side of the above equation is the vorticity transport rate due to diffusion, the second term is the vorticity transport rate due to fluidity gradient, and the third term is the vorticity generation rate due to the contribution of elastic stresses from micellar solutions. Figure~\ref{vorticityTansport_Re62} shows the VTR distribution at different time instances, both for Newtonian fluids at $Re = 62$ and micellar solutions at $Re = 40$ and 62. Irrespective of the fluid type, VTR is prominent in the shear layer on the front face of the cylinder, then decays progressively downstream. The noticeable differences between the two fluids are as follows: The VTR is more concentrated and less diffused in micellar solutions than in Newtonian fluids, both near the cylinder and downstream of it. A higher VTR is observed in the thinned regions along the edges of the downstream wakes for micellar solutions at $Re = 40$ (see first column of figure~\ref{vorticityTansport_Re62}). As the Reynolds number increases to 62, the vorticity transport rate becomes pronounced near the cylinder in the downstream region (see second column of figure~\ref{vorticityTansport_Re62}), whereas it is almost absent in Newtonian fluids (see the third column in figure~\ref{vorticityTansport_Re62}). This concentration of VTR around the cylinder and along the edges of the recirculation regions is due to micelles breaking in these regions under strong shearing and extensional flows, respectively. As a result, fluidity increases in these regions, leading to reduced VTR diffusion (the first term in the VTR equation). Furthermore, strong gradients in both fluidity $(\bm{\nabla} \phi)$ and vorticity $(\bm{\nabla} \omega)$ are present in these regions (the second term in the VTR equation), which again result in a decrease in the overall VTR diffusion. Furthermore, elastic stresses will also contribute to VTR generation in these regions for micellar solutions (the third term in the VTR equation). All these factors ultimately lead to a higher vorticity transport rate in micellar solutions, which is again confined to relatively thinner regions than in Newtonian fluids under otherwise identical conditions.  

\subsection{Surface integral parameters} 

\begin{figure}
    \centering
    \includegraphics[width=14cm]{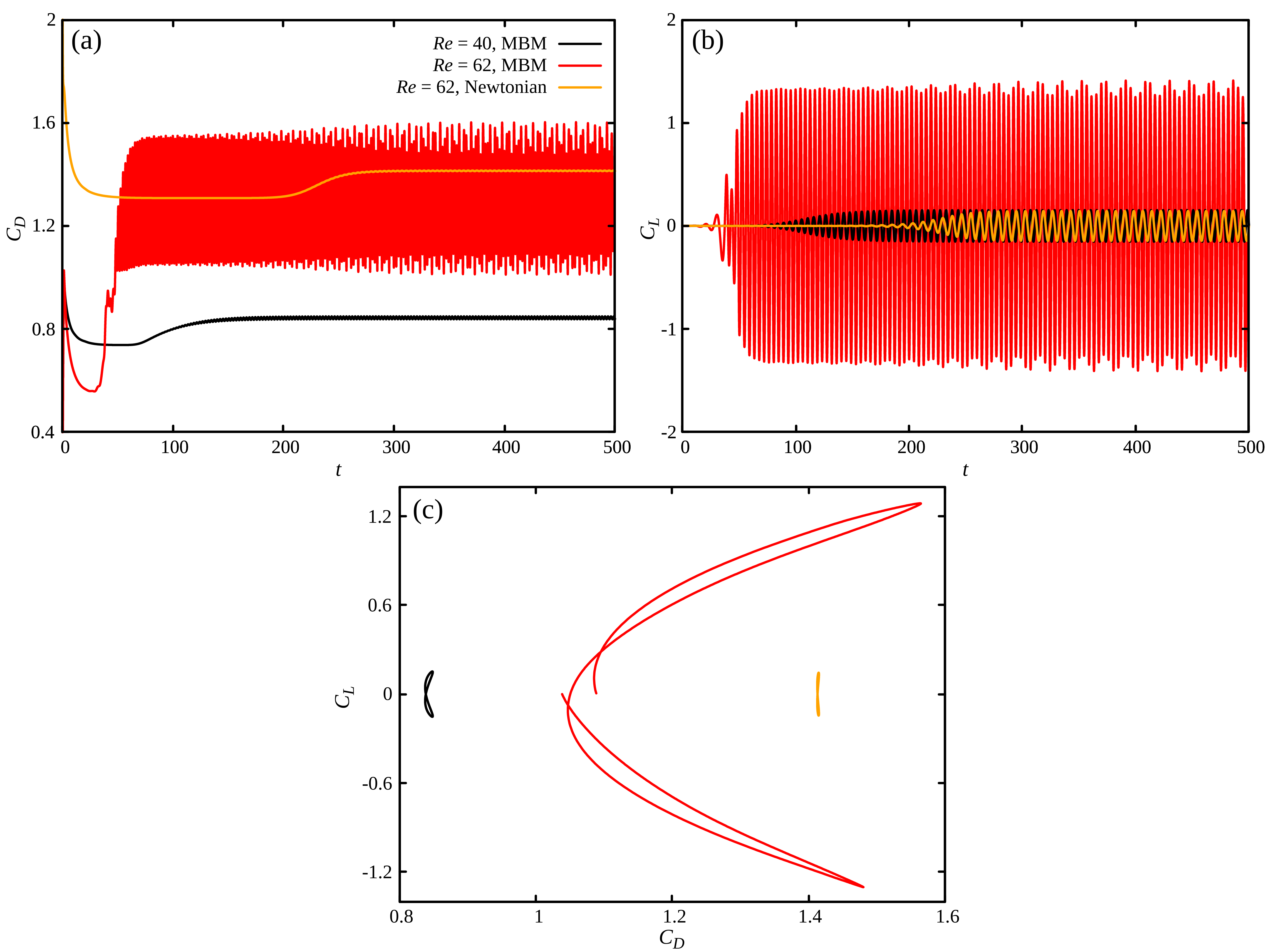}
    \caption{Temporal variation of drag coefficient (a), and lift coefficient (b) at different Reynolds numbers. Sub-figure~(c) shows the phase diagram of the drag and lift coefficient.}
    \label{Cd_Cl}
\end{figure}

The surface integral parameters of engineering importance, such as drag and lift coefficients, are often required for various design calculations involving flow past an obstacle. These are calculated as follows

\begin{equation}
    C_{D} = \frac{1}{\frac{1}{2} \rho U_{\infty}^{2} D_0} \int_{S} (( -p\bm{\delta} + \bm{\tau}) \cdot \bm{n}) \cdot e_{x} \,\text{dS}
\end{equation}

\begin{equation}
    C_{L} = \frac{1}{\frac{1}{2} \rho U_{\infty}^{2} D_0} \int_{S} ((-p\bm{\delta} + \bm{\tau}) \cdot \bm{n}) \cdot e_{y} \,\text{dS}
\end{equation}

The drag $(C_D)$ and lift $(C_L)$ coefficients measure the non-dimensional forces per unit length (consisting of pressure, friction, and elastic forces) acting on the cylinder in the flow direction and in the normal direction, respectively. Sub-figures~\ref{Cd_Cl}(a) and (b) show the temporal variation of the drag and lift coefficients both for Newtonian fluids and micellar solutions at three different Reynolds numbers, namely, 30, 40, and 62. At $Re = 30$, the drag coefficient reaches a plateau value with time, and the lift coefficient is zero both for Newtonian fluids and micellar solutions, again confirming the presence of a steady and symmetric flow field at this condition. In contrast, at $Re = 40$, the micellar solutions exhibit periodic fluctuations in both drag and lift coefficients, suggesting a transition from a steady to an unsteady flow state. However, the corresponding Newtonian case again exhibits a constant drag coefficient and a zero lift coefficient, still indicating the presence of a steady state in the flow field. On the other hand, micellar solution shows a quasi-periodic fluctuation both in drag and lift coefficients, whereas Newtonian fluids show a regular periodic fluctuation in both quantities at $Re = 62$. These temporal variations in drag and lift coefficients provide useful information about the flow state in the present system, as do the temporal variations in velocity components at a probe location discussed in the preceding sub-section. Sub-figure~\ref{Cd_Cl}(c) displays the phase diagram of drag and lift coefficients. At $Re = 40$ for micellar solutions and at $Re=62$ for Newtonian fluids, the phase diagram is closed, once again suggesting the presence of a regular periodic motion at these conditions for the respective fluids. However, it is non-closed for micellar solutions at $Re = 62$, and the enclosed area increases more. This lack of closure in the phase trajectory indicates that the flow does not evolve on a simple limit cycle. Instead, multiple interacting time scales are present, leading to quasi-periodic dynamics in micellar solutions under this flow condition. On the other hand, an increase in the enclosed area of the phase diagram indicates enhanced nonlinear coupling between lift and drag forces, reflecting stronger vortex-induced energy exchange and increased oscillation amplitude, suggesting a transition toward more energetic and dynamically active flow behaviour in micellar solutions.           

\begin{figure}
    \centering
    \includegraphics[width=14cm]{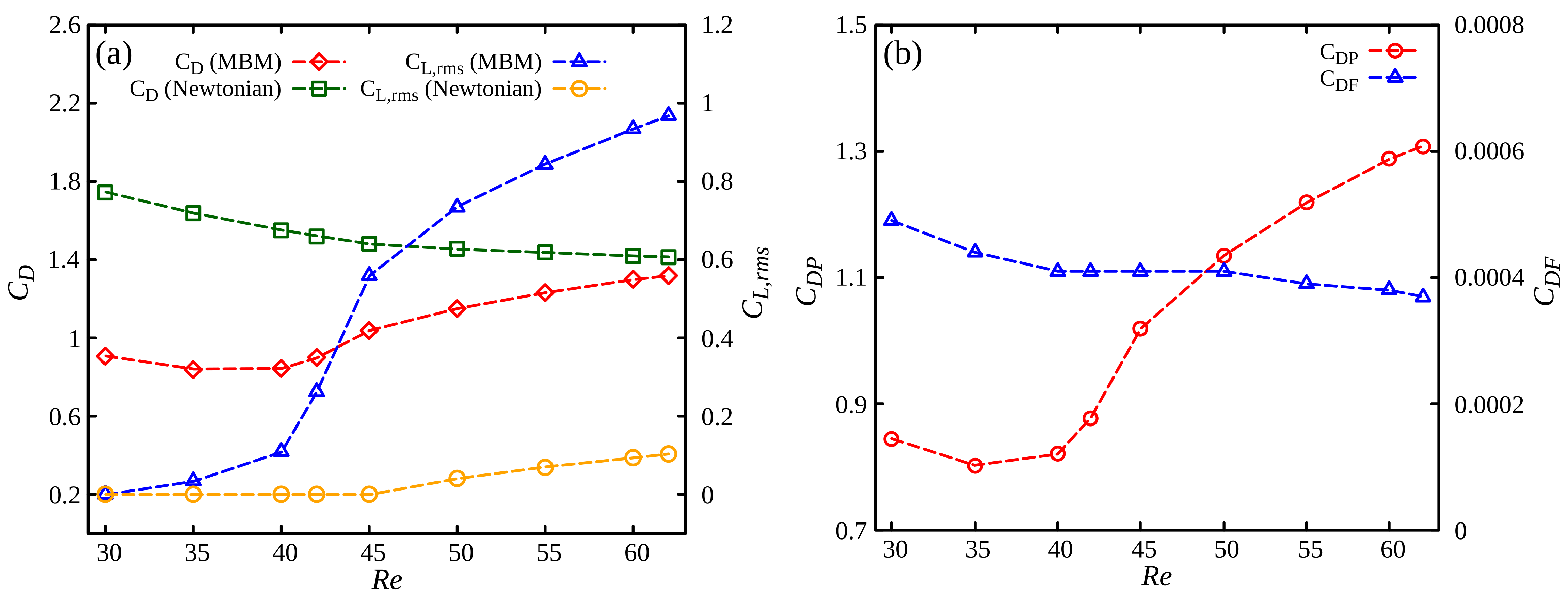}
    \caption{Variation of drag and lift coefficients with Reynolds number, both for Newtonian fluids and micellar solutions (a). The component variation of drag coefficient, namely, pressure ($C_{DP}$) and friction ($C_{DF}$), for micellar solutions (b).}
    \label{DragAndLift_Time-Averaged}
\end{figure}

\begin{figure}
    \centering
    \includegraphics[width=13cm]{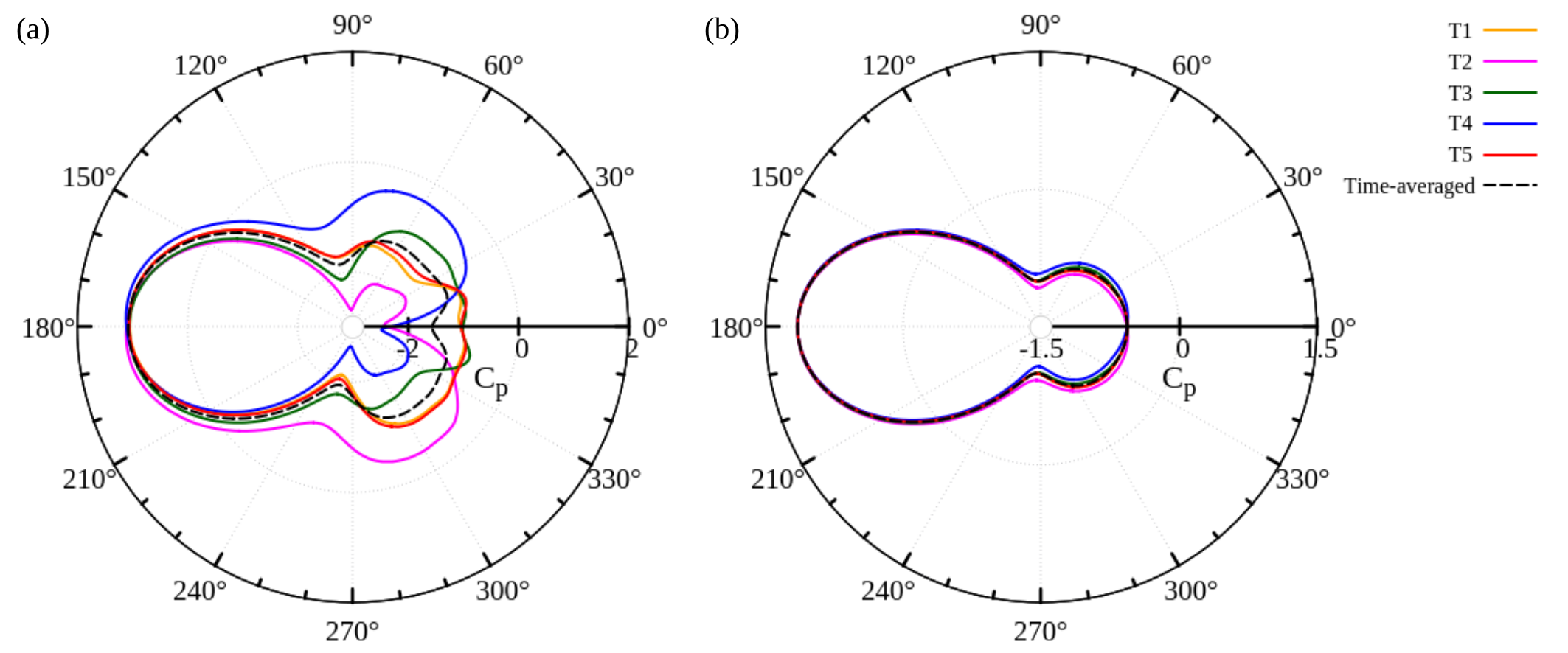}
    \caption{Variation of pressure coefficient $(C_p)$ along the surface  of the cylinder for micellar solutions (a) and Newtonian fluids (b) at $Re = 62$. Here, the angle is measured from the cylinder's rear end in anti-clockwise direction.}
    \label{PressureCoefficient_MBM_Re62}
\end{figure}

However, from a practical application viewpoint, it is the steady and time-averaged (in the case of unsteady flows) drag and lift coefficients that are more useful, and therefore, figure~\ref{DragAndLift_Time-Averaged} shows their variations with Reynolds number both for Newtonian fluids and micellar solutions. For Newtonian fluids, the drag and lift coefficients exhibit the well-known decreasing and increasing trends with Reynolds number, respectively. As the Reynolds number increases, the hydrodynamic boundary layer $(\delta_h)$ becomes thinner, scaled as $\delta_h \sim \frac{1}{\sqrt{Re}}$, where the viscous forces act. As a result, the contribution of viscous drag decreases with increasing Reynolds number, thereby reducing the total drag, as shown in sub-figure~\ref{DragAndLift_Time-Averaged}(a). In contrast, the variation of the drag coefficient with the Reynolds number for micellar solutions exhibits a different trend. It first decreases slightly with $Re$, then increases gradually rather than decreasing further. This increase in drag in micellar solutions is explained as follows: the frictional drag contribution is much lower in micellar solutions than in Newtonian fluids. This is due to high fluidity in the vicinity of the cylinder, resulting from the breakage of micelles, which lowers the effective viscosity and, consequently, the frictional drag. Therefore, the drag in micellar solutions is mostly contributed by the pressure component, which increases sharply with Reynolds number, as can be seen from sub-figure~\ref{DragAndLift_Time-Averaged}(b). This increase in pressure drag with Reynolds number in micellar solutions resulted from earlier boundary-layer separation in these solutions compared to Newtonian fluids, which, in turn, increased the pressure difference and hence the pressure drag. To explicitly show this, figure~\ref{PressureCoefficient_MBM_Re62} shows the plot of non-dimensional pressure coefficient $\left(C_p = \frac{p}{\frac{1}{2} \rho U_{\infty}^2}\right)$ on the cylinder surface at different time instants, both for Newtonian and micellar solutions at a particular value of $Re = 62$. First of all, regardless of the fluid type, the pressure is high at the cylinder's front stagnation point, where fluid parcels come to rest, converting their kinetic energy into pressure. For Newtonian fluids, it gradually decreases along the cylinder surface, reaching a minimum at around $\theta = 90^{\circ}$, then increases somewhat in the downstream rear stagnation region, thereby leading to some pressure recovery in this region (sub-figure~\ref{PressureCoefficient_MBM_Re62}(b)). Furthermore, the pressure distribution is symmetric across the upper and lower halves of the cylinder throughout the cycle, confirming once again the presence of regular periodic flow in Newtonian fluids under this condition. On the other hand, in micellar solutions, the pressure variation is much larger along the cylinder surface than in Newtonian fluids (sub-figure~\ref{PressureCoefficient_MBM_Re62}(a)). At certain times, for instance at T4, the pressure at the rear stagnation point decreases more sharply because of the formation of an almost stagnation zone due to the presence of small recirculation wakes (see figure~\ref{Stream_Re62}). This ultimately causes a greater pressure difference in the streamwise direction, thereby significantly increasing the pressure drag in micellar solutions.                   

On the other hand, the lift coefficient increases for both fluids (sub-figure~\ref{DragAndLift_Time-Averaged}(a)). However, it is much larger, and the Reynolds-number dependence is much steeper in micellar solutions than in Newtonian fluids. For instance, at $Re = 62$, the lift coefficient value is almost 9 times larger for micellar solutions than for Newtonian fluids. This is again attributed to a large pressure imbalance between the upper and lower halves of the cylinder, originating in micellar solutions. For instance, at times T2 and T4, one can observe a large vertical asymmetry in the pressure distribution across the cylinder's upper and lower halves during the flow of micellar solutions. This, in turn, results in a larger lift coefficient in micellar solutions than in Newtonian fluids, where a nearly symmetric pressure distribution is observed.

\subsection{Dynamic mode decomposition analysis}

\begin{figure}
    \centering
    \includegraphics[width=13cm]{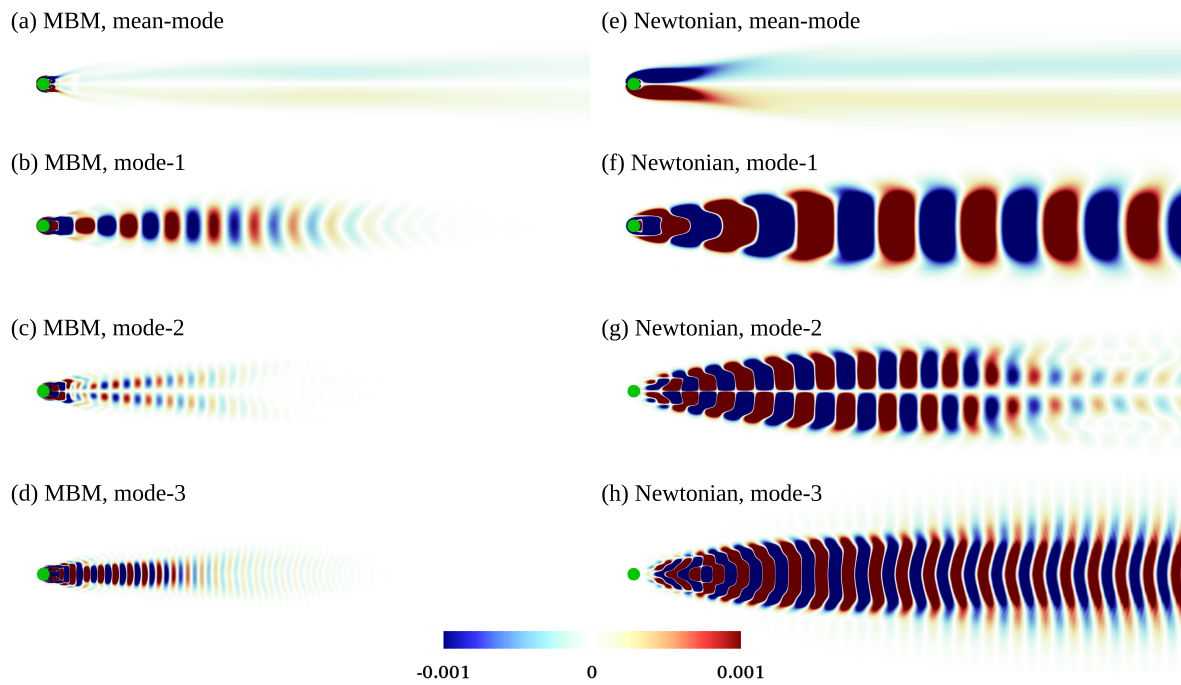}
    \caption{Visual representation of DMD modes of the vorticity field for micellar solutions (first column) and Newtonian fluids (second column) at $Re = 62$.}
    \label{DMD}
\end{figure}

The dynamic mode decomposition (DMD) analysis is performed using the algorithm originally proposed by Schmid~\citep{schmid2011}, and further implemented in our previous studies~\citep{hamid2022dynamic,hamid2023significant} at a fixed value of $Re = 62$, to further analyse the differences in the flow field of micellar solutions and Newtonian fluids under the same flow conditions. In the present work, a sequence of temporally equispaced snapshots of the vorticity field is collected in the statistically stationary unsteady regime ($t $). A total of $N = 200$ and 800 snapshots are sampled at a uniform time interval of $\Delta t = 0.172$ and 0.086 for Newtonian and micellar solutions, respectively. Each snapshot is vectorised and arranged column-wise to construct the data matrix $S_{1}^{N} = \{s_j\}_{j=1}^{N}$, where $s_j \in \mathbb{R}^m$ represents the discretised vorticity field at time $t_j$. DMD assumes that the temporal evolution of the flow can be approximated by a linear operator $\mathbf{M}$ that maps consecutive snapshots, i.e., $s_{j+1} = \mathbf{M} s_j$. Accordingly, the dataset can be partitioned into two matrices,
\begin{equation}
S_{1}^{N-1} = [s_1, s_2, \dots, s_{N-1}], \quad
S_{2}^{N} = [s_2, s_3, \dots, s_N],
\end{equation}
such that $S_{2}^{N} \approx \mathbf{M} S_{1}^{N-1}$. In practice, a reduced-order approximation $\mathbf{C}$ of $\mathbf{M}$ is computed by minimising the residual $\mathbf{r} = S_{2}^{N} - \mathbf{C} S_{1}^{N-1}$ in a least-squares sense using singular value decomposition (SVD), yielding a low-dimensional representation of the dominant flow dynamics.

The eigendecomposition of $\mathbf{C}$ provides the DMD eigenvalues (Ritz values, $\lambda_j$) and eigenvectors, from which the corresponding DMD modes ($\phi_j$) are reconstructed. The DMD modes represent the coherent spatial structures of the flow, while the Ritz values encode their temporal behaviour. Specifically, each eigenvalue $\lambda_j$ is related to the continuous-time growth/decay rate and oscillation frequency of the corresponding mode through:
\begin{equation}
\sigma_j = \frac{\ln|\lambda_j|}{\Delta t}, \qquad
f_j = \frac{\operatorname{Im}(\ln \lambda_j)}{2\pi \Delta t},
\end{equation}
where $\sigma_j$ denotes the modal growth (or decay) rate and $f_j$ represents the frequency. The relative importance of each mode is quantified through its amplitude $b_j$, which is computed by projecting the initial snapshot onto the DMD modal basis as $\mathbf{b} = \mathbf{\Phi}^{\dagger} s_1$, where $\mathbf{\Phi}$ is the matrix of DMD modes and $\mathbf{\Phi}^{\dagger}$ denotes its Moore--Penrose pseudoinverse. The magnitude of $b_j$ provides a measure of the energy contribution of each mode, allowing identification of the dominant coherent structures governing the flow dynamics.

The DMD analysis of the vorticity field clearly reveals distinct differences in the flow physics between the micellar solution and the Newtonian case (figure~\ref{DMD}). The mean mode represents the time-averaged wake structure, in which the micellar solution exhibits significantly concentrated vortex structures near the cylinder, which are again confined to a thin region around the cylinder. In contrast, it is more dispersed and widened in Newtonian fluids. A significant difference is also observed in higher modes. For instance, in modes 2 and 3, the vortical structures are away from the cylinder surface in Newtonian fluids, whereas they are concentrated near the cylinder surface in micellar solutions. Once again, the vortical structures in the region downstream of the cylinder are confined to a smaller region in micellar solutions than in Newtonian fluids.  

\begin{figure}
    \centering
    \includegraphics[width=13cm]{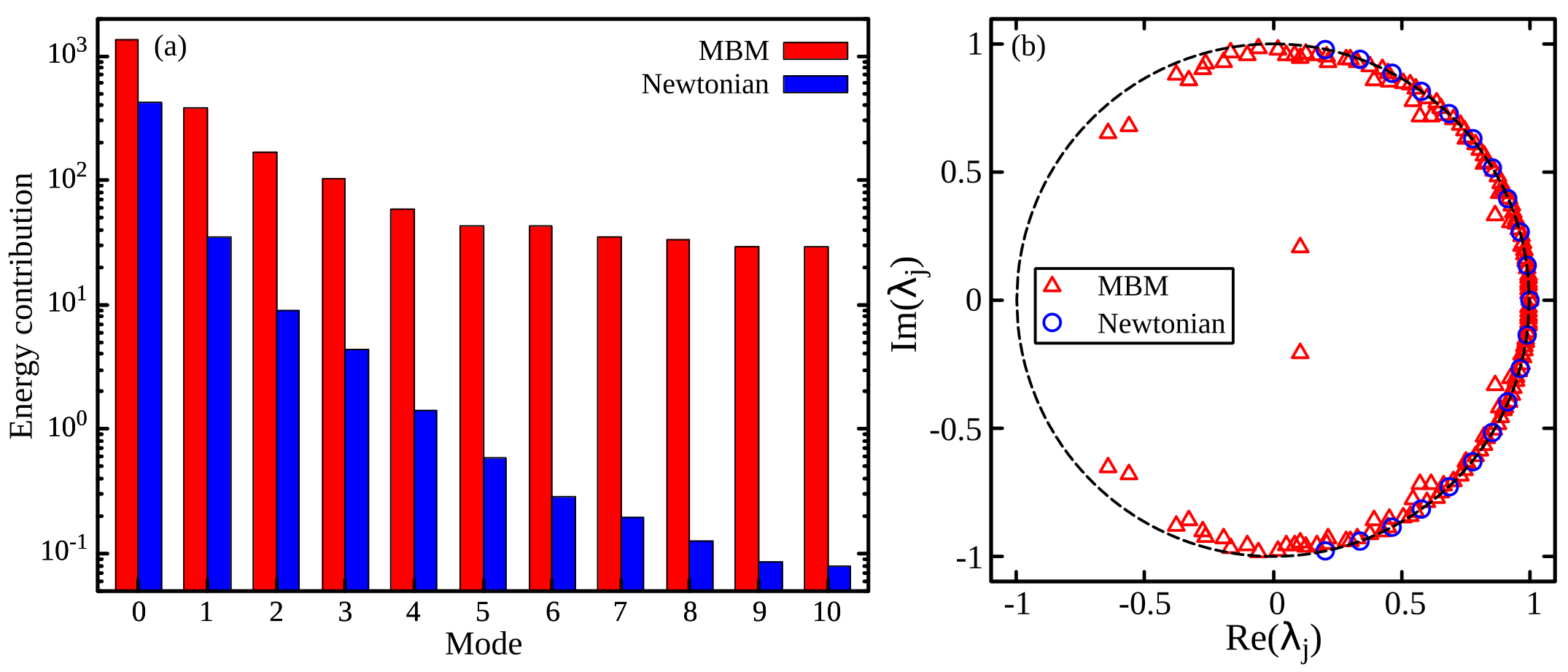}
    \caption{Energy contribution of first few DMD modes (a), and Ritz values, $\lambda_j$ (b) for both micellar solutions and Newtonian fluids at $Re = 62$. Here, $0^{\text{th}}$ mode implies the mean DMD mode in sub-figure (a).}
    \label{energyContribution}
\end{figure}

The energy contribution plot shows that the micellar solution retains significantly more energy across higher modes than the Newtonian fluid (sub-figure~\ref{energyContribution}(a)). In Newtonian fluids, energy decays sharply with mode number, indicating that most of the dynamics are captured by the first few dominant modes (low-dimensional behaviour). However, in the micellar solution, energy decays much more slowly, implying a broader distribution of energy across modes and hence a higher-dimensional, more complex flow. This is a direct consequence of spatial and temporal changes in fluidity, resulting from the breaking and reforming of micelles, which introduces additional time and length scales, leading to sustained fluctuations even at smaller scales.

The Ritz value distribution clearly distinguishes micellar solutions from Newtonian fluids in terms of spectral spread and dynamical richness (see sub-figure~\ref{energyContribution}(b)). In the Newtonian case, the eigenvalues are tightly clustered along the unit circle, particularly in the right half-plane, indicating nearly neutrally stable, periodic vortex shedding with minimal damping or growth, consistent with a dominant, coherent limit-cycle behaviour. In contrast, the micellar solution exhibits a noticeably broader and more scattered distribution. While many modes still align near the unit circle, corresponding to non-decaying oscillatory modes likewise seen in Newtonian fluids, several others deviate inward and even appear within the unit circle. This spread reflects increased modal damping and the presence of multiple time scales introduced by micellar dynamics, such as continuous breaking and reformation of structures. The deviation from a sharp unit-circle clustering suggests that, unlike the Newtonian case, the micellar solution is less purely periodic and contains transient, dissipative, and possibly intermittent dynamics that alter the stability and energy transfer mechanisms in the flow.

\section{Probable mechanism for early flow transition \label{mechanism_flow_transition}}

\begin{figure}
    \centering
    \includegraphics[width=13cm]{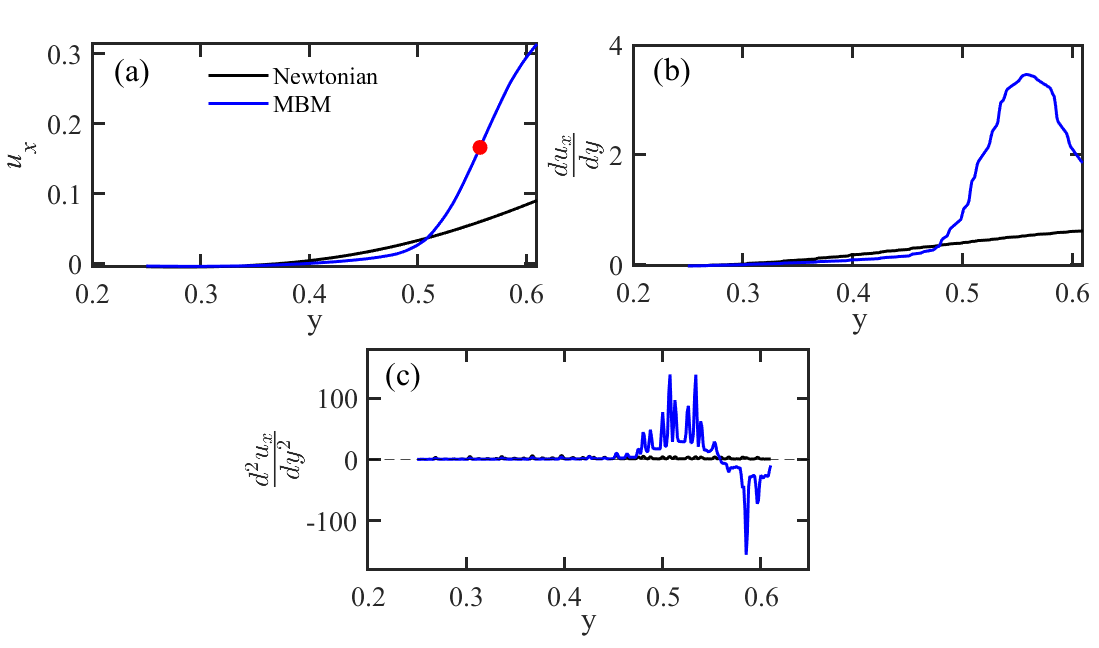}
    \caption{Variation of streamwise velocity component along a vertical line starting at $(x_{1}, y_{1} = 0.5, 0.25)$ and ending at $(x_{2}, y_{2} = 0.5, 0.61)$ downstream of the cylinder passing through the shear layer both for Newtonian fluids and micellar solutions at $Re = 40$ (a). Spanwise variation of the first (b) and second (c) order derivatives of the streamwise velocity component with respect to the spanwise direction.}
    \label{KHplot}
\end{figure}

All the aforementioned results and discussion indicate that an early transition from steady to unsteady flow occurs in micellar solutions than in Newtonian fluids. A probable mechanism for this early flow transition is presented here. First, we can see from sub-figure~\ref{Fig:Stream_Re40}(f) that the velocity changes sharply across the shear layer in micellar solutions, whereas in Newtonian fluids it changes gradually, as shown in sub-figure~\ref{Fig:Stream_Re40}(g). This sudden change in velocity occurs because of the presence of a high-fluidity region in the shear layer, resulting from micelle destruction (sub-figure~\ref{Fluidity_Re40}(f)). This might have triggered the Kelvin-Helmholtz (KH) instability in this region, potentially destabilising the flow even at lower Reynolds numbers. Figure~\ref{KHplot} shows the plot of the streamwise velocity component along a vertical line passing through the shear layer, along with its first and second derivatives with respect to the spanwise distance. It can be seen that the second derivative (sub-figure~\ref{KHplot}(c)) changes its sign across a point within the shear layer region (as marked by a red circle in sub-figure~\ref{KHplot}(a)), thereby suggesting the presence of an inflection point around which the KH instability could have triggered. 

\begin{figure}
    \centering
    \includegraphics[width=13cm]{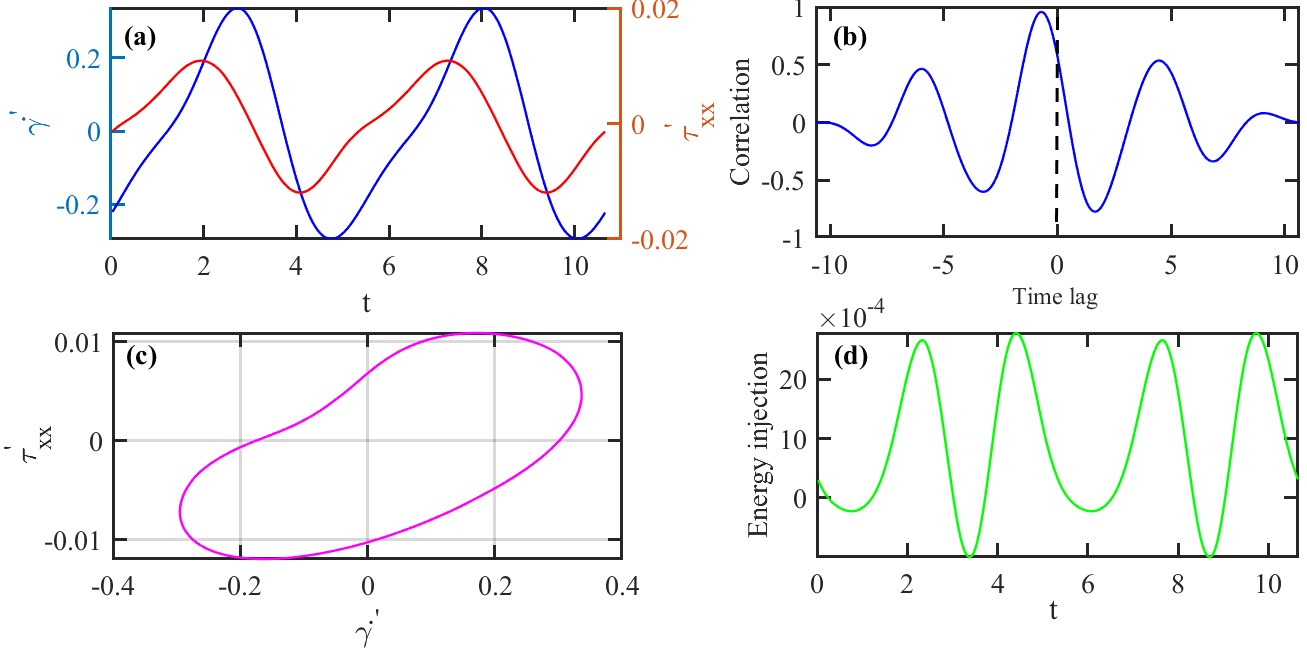}
    \caption{(a) Temporal variation of the streamwise normal stress component $\tau_{xx}^{'}$ and shear rate magnitude $\dot{\gamma}^{'}$ fluctuation at the same probe location situated on the vertical line where an inflection point is observed in sub-figure~\ref{KHplot}(a). The corresponding (b) cross-correlation and (c) phase or Lissajous plot between the two. (d) The energy injection over time into the flow field.}
    \label{cross}
\end{figure}

Figure~\ref{cross} clearly demonstrates that the acceleration of shear-layer instability in the micellar solutions originate from the phase-lagged coupling between elastic stress and shear rate fluctuations, which leads to net energy injection into the flow. In sub-figure~\ref{cross}(a), the streamwise normal stress fluctuation $\tau_{xx}^{'}( = \tau_{xx} - <\tau_{xx}>)$ is not in phase with the shear rate fluctuation $\dot{\gamma}^{'} (= \dot{\gamma} - <\dot{\gamma}^{'}>)$, calculated at the same inflection point within the shear layer, as shown in sub-figure~\ref{KHplot}(a), indicating a delayed viscoelastic response arising from the underlying thixotropic microstructure dynamics. This is quantitatively confirmed by the shifted peak in the cross-correlation plot in sub-figure~\ref{cross}(b), which establishes a finite time lag between deformation and stress generation. The Lissajous curve in sub-figure~\ref{cross}(c) forms a hysteresis loop, revealing that the stress-strain interaction is not purely dissipative but involves a cyclic exchange of energy, with a non-zero loop area indicating net work done by the elastic stresses. Consequently, as shown in sub-figure~\ref{cross}(d), there are distinct bursts of positive energy injection into the flow when stress and shear align constructively. This delayed yet reinforcing feedback mechanism allows elastic stresses to amplify velocity gradients rather than damp them, effectively acting as a destabilising force. As a result, perturbations in the shear layer grow more rapidly, leading to an earlier onset and enhanced development of shear-layer instability in the micellar solutions compared to a Newtonian counterpart.

\begin{figure}
    \centering
    \includegraphics[width=10cm]{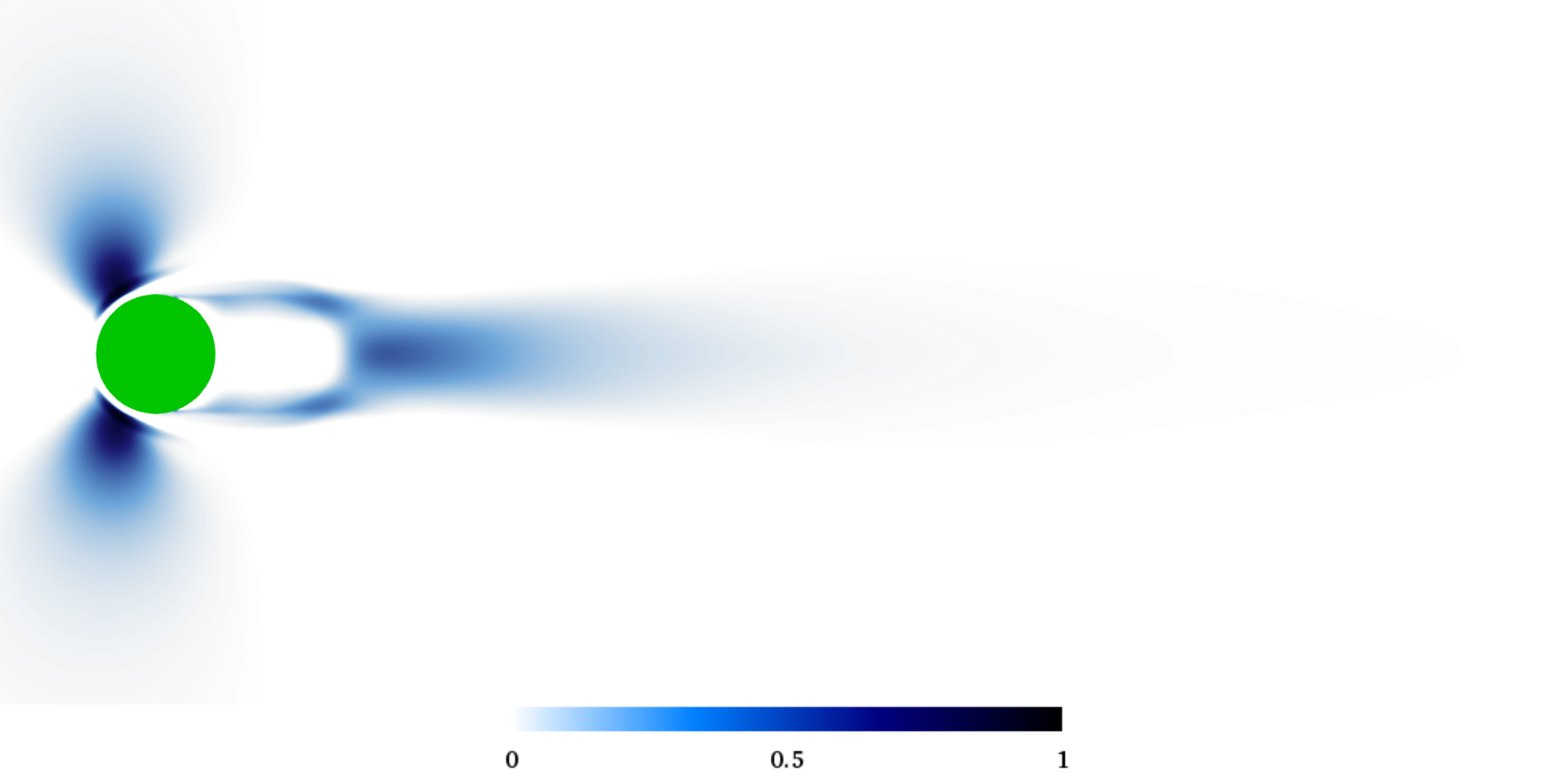}
    \caption{Surface distribution of time-averaged streamwise elastic stress component $\tau_{xx}$ at $Re = 40$ for micellar solutions.}
    \label{stress}
\end{figure}

This is in contrast to viscoelastic polymer solutions, wherein the shear-layer instability is delayed to a higher Reynolds number~\citep{hamid2022dynamic,sahin2004effects}. In these solutions, where the breaking and reformation dynamics are absent, a strand of high streamwise-normal elastic stress was observed in the shear layers due to the high alignment and stretching of polymer molecules in this region, which inhibits the corresponding KH instability~\citep{hamid2022dynamic}. However, in the present case of micellar solutions, micelles are broken in the shear-layer, resulting in a decrease of elastic stresses in this region (see figure~\ref{stress}), which in turn can not suppress the instability. In addition, a phase lag between stress and deformation is present, as mentioned above, thereby preventing the generation of higher stresses when deformation is higher to suppress the instability. Therefore, lower values of elastic stress and its phase lag with deformation ultimately accelerate instability in micellar solutions.

\section{Conclusions}

In conclusion, the present study establishes that the flow behaviour of micellar solutions past a circular cylinder is fundamentally governed by the strong coupling between flow kinematics and microstructural kinetics, as captured by the MBM model. By incorporating both viscoelastic stress evolution and reversible micellar breakage-reformation dynamics, the simulations, parameterised using experimentally fitted data for the EHAC-NaSal system, demonstrate that micellar fluids exhibit markedly richer and more complex dynamics than Newtonian fluids, even at low-to-intermediate Reynolds numbers. A key outcome is the earlier transition from steady to unsteady flow, which is traced to the interplay of relatively low but dynamically evolving elastic stresses, phase-lagged stress-strain response, and thixotropic restructuring. Unlike purely viscous fluids, the delayed elastic response introduces a finite phase difference between stress and deformation, enabling intermittent positive energy transfer to the flow, accelerating shear-layer instability and promoting premature transition. As the Reynolds number increases, the flow departs from classical periodic vortex shedding and exhibits quasi-periodic dynamics, reflecting the additional temporal scales introduced by micellar kinetics. This behaviour is also manifested in non-monotonic drag variation, increased lift fluctuations, and an increase in the Strouhal number, all of which indicate stronger and more complex unsteady forcing. The wake structure shows a regime-dependent modification: micellar fluids produce larger recirculation zones in the steady regime, while in the unsteady regime the wake becomes more compact and spatially localised, with vorticity confined to thin shear layers and intensified transport near the cylinder. Data-driven DMD analysis further reveals the coexistence of decaying and self-sustained modes, highlighting competing stabilising and destabilising mechanisms introduced by microstructural evolution, in contrast to the purely self-sustained dynamics of Newtonian flows. Overall, these findings underscore that microstructural kinetics, phase lag, and elastic energy exchange collectively dictate the onset of instability, vortex dynamics, and force characteristics in micellar flows. The ability of such fluids to trigger early transition provides new physical insight and offers opportunities to actively tune flow behaviour in micellar and other structured fluid systems. 

\section*{Acknowledgments}
We acknowledge the National Supercomputing Mission (NSM) for providing computing resources of ‘PARAM Smriti’ at NABI, Mohali (accessed by CS) and ‘PARAM Himalaya’ at IIT Mandi (accessed by AC), which are implemented by C-DAC and supported by the Ministry of Electronics and Information Technology (MeitY) and Department of Science and Technology (DST), Government of India. CS acknowledges financial support from the Anusandhan National Research Foundation (ANRF), Government of India, through a core research grant (CRG/2023/001908), and AC thanks the Ministry of Education, Government of India, for financial support from the PMRF (Cycle-9).

\section*{Supplementary Material}
Supplementary video on vortex dynamics in micellar solutions and Newtonian fluids at $Re = 62$. 

\section*{Declaration of Interests}
The authors report no conflict of interest.
\appendix

\section{\label{App:A}Determination of the MBM constitutive model parameters from the experimental rheological data on WLM solutions}

\begin{figure}
    \centering
    \includegraphics[width=13cm]{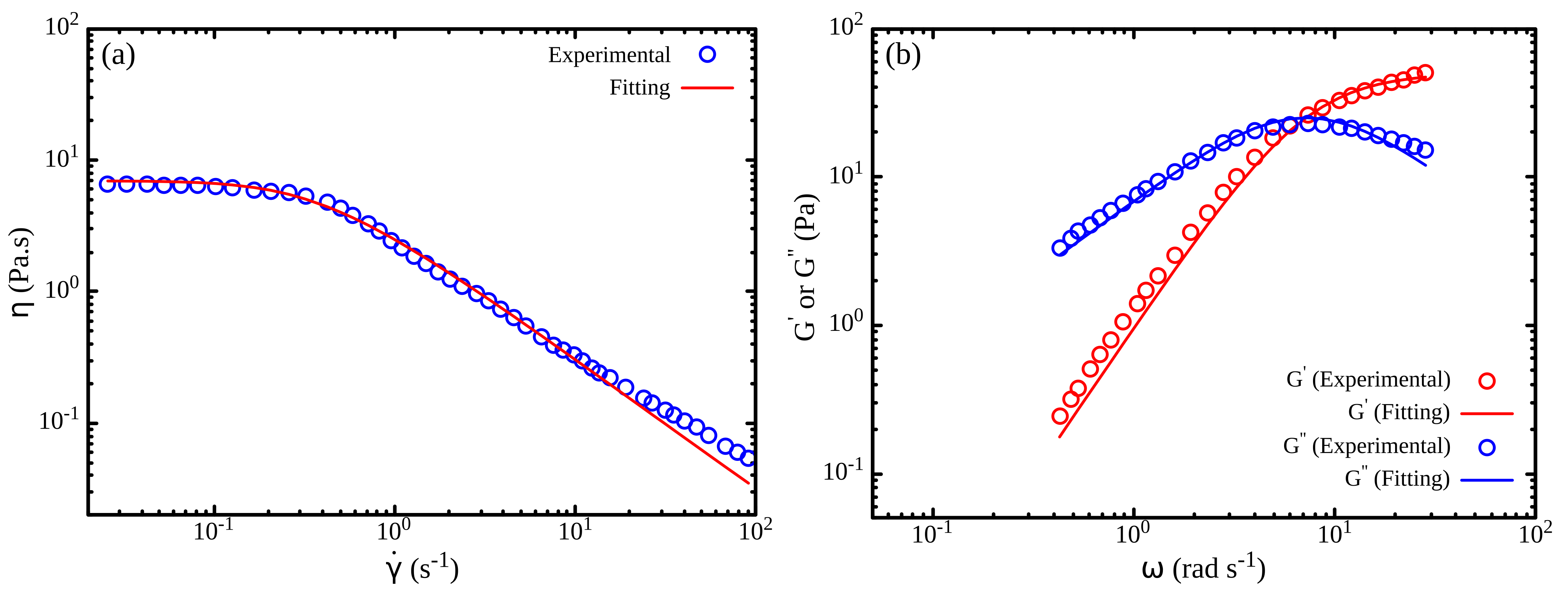}
    \caption{Fitting curve of the present implemented model with that of the experimental data of Raghavan and Kaler~\citep{raghavan2001highly}. Steady-shear viscosity (a) and storage ($\text{G}'$) and loss ($\text{G}''$) moduli (b). The experimental results shown here are for NaSal/EHAC solution with a molar ratio of 0.65 at $25~^\circ C$.}
    \label{fitting}
\end{figure}

The MBM model parameters used in this study are obtained by fitting the experimental results of steady shear and small amplitude oscillatory shear (SAOS) test results provided by Raghavan and Kaler~\citep{raghavan2001highly} for erucyl bis(hydroxyethyl)methylammonium chloride (EHAC) and sodium salicylate (NaSal) micellar system with a molar ratio of 0.65 at 25$^\circ$C. For a homogeneous simple shear flow with a shear rate $\dot{\gamma}  (t)$, the MBM model can be written in matrix form as follows:

\begin{equation}
    \begin{split}
        \begin{bmatrix}
         \tau_{p,xx} & \tau_{p,xy} & 0\\
         \tau_{p,yx} & \tau_{p,yy} & 0\\ 
         0& 0 & \tau_{p,zz}\\
        \end{bmatrix} + \frac{1}{G_{0}\phi} \frac{d}{dt}\begin{bmatrix}
         \tau_{p,xx} & \tau_{p,xy} & 0\\
         \tau_{p,yx} & \tau_{p,yy} & 0\\ 
         0& 0 & \tau_{p,zz}\\
        \end{bmatrix} - \frac{1}{G_{0}\phi}\lambda \begin{bmatrix}
         0 & \dot{\gamma} & 0\\
         0 & 0 & 0\\ 
         0& 0 & 0\\
        \end{bmatrix} \begin{bmatrix}
         \tau_{p,xx} & \tau_{p,xy} & 0\\
         \tau_{p,yx} & \tau_{p,yy} & 0\\ 
         0& 0 & \tau_{p,zz}\\
        \end{bmatrix} - \\
       \frac{1}{G_{0}\phi} \lambda \begin{bmatrix}
         \tau_{p,xx} & \tau_{p,xy} & 0\\
         \tau_{p,yx} & \tau_{p,yy} & 0\\ 
         0& 0 & \tau_{p,zz}\\
        \end{bmatrix} \begin{bmatrix}
         0 & 0 & 0\\
         \dot{\gamma} & 0 & 0\\ 
         0& 0 & 0\\
        \end{bmatrix} = \frac{1}{\phi} \begin{bmatrix}
         0 & \dot{\gamma} & 0\\
         \dot{\gamma} & 0 & 0\\ 
         0& 0 & 0\\
        \end{bmatrix} 
\end{split}
\end{equation}

Therefore, the shearing component of the stress tensor becomes

\begin{equation}
    \tau_{p,xy} + \frac{1}{G_{0}\phi} \frac{d \tau_{p,xy}}{dt} = \frac{\dot{{\gamma}}}{\phi}
\end{equation}

Furthermore, the fluidity equation under shearing flow becomes

\begin{equation}
    \frac{\partial \phi}{\partial t} = \frac{1}{\lambda} \left(\phi_{0} - \phi\right) + \frac{K_{0}}{\eta_{\infty}}~\dot{\gamma}~\tau_{p,xy}
    \label{fluid_SS}
\end{equation}

Note that the gradients of both the stress tensor and fluidity become zero due to the homogeneity condition, i.e., $\bm{\nabla} \bm{\tau_p} = \bm{\nabla} \phi = 0$. For steady shear flow, i.e., $\dot{\gamma}(t) = \dot{\gamma}_0$, the above two equations become

\begin{equation}
    \tau_{p,xy} = \frac{\dot{\gamma}}{\phi}
\end{equation}

\begin{equation}
    \frac{1}{\lambda} \left(\phi_{0} - \phi\right) + \frac{K_{0}}{\eta_{\infty}}~\dot{\gamma}~\tau_{p,xy} = 0
\end{equation}

After substituting the expression for $\tau_{p,xy}$ in the fluidity equation, we get a quadratic equation for $\phi$ as follows

\begin{equation}
    \phi^{2} - \phi_{0} \phi - \frac{K_{0}}{\eta_{\infty}}\lambda \dot{\gamma}^{2} = 0
\end{equation}

The solution of the above equation gives $\phi = \frac{\phi_{0}}{2}\left[1 + \sqrt{1+4 \lambda \frac{K_{0}}{\eta_{\infty}} \frac{\dot{\gamma}^{2}}{\phi_{0}^{2}}}\right]$. The total shear stress then becomes

\begin{equation}
    \tau_{xy} = \tau_{s,xy} + \tau_{p,xy} = \eta_{s} \dot{\gamma}_{0} + \frac{\dot{\gamma}_{0}}{\frac{\phi_{0}}{2}\left[1 + \sqrt{1+4 \lambda \frac{K_{0}}{\eta_{\infty}} \frac{\dot{\gamma}^{2}}{\phi_{0}^{2}}}\right]}
\end{equation}

The apparent viscosity becomes

\begin{equation}
    \eta = \eta_{s} + \frac{2 \eta_{0}}{\left[1 + \sqrt{1+4 \lambda \frac{K_{0}}{\eta_{\infty}} \dot{\gamma}^{2}{\eta_{0}^{2}}}\right]}
\label{app_vis}
\end{equation}

On the other hand, for a SAOS flow, after substituting $\dot{\gamma}(t) = \dot{\gamma}_{0} \text{cos}\,\omega t$ and assuming $\phi = \phi_{0}$ in equation~\ref{fluid_SS}, we get 

\begin{equation}
    \tau_{p,xy} + \frac{1}{G_{0}\phi_{0}} \frac{d \tau_{p,xy}}{dt} = \frac{\dot{\gamma}_{0}\, \text{cos}\,\omega t}{\phi_{0}}
\label{SAOS}
\end{equation}

\begin{equation}
    \tau_{s,xy} = \eta_{s} \dot{\gamma}_{0}\, \text{cos}\, \omega t
\end{equation}

Equation~\ref{SAOS} is a first-order ordinary differential equation that can be solved using the integrating factor (IF) as $e^{G_{0}\phi_{0}t}$. After performing the rearrangement, the total extra stress tensor can be written as

\begin{equation}
    \tau_{xy} = \tau_{s,xy}+\tau_{p,xy} = \frac{\dot{\gamma}}{\omega} \left[G'(\omega) \text{sin} (\omega t) + G''(\omega) \text{sin}(\omega t)\right]
\end{equation}

where the storage $(G')$ and loss $(G'')$ moduli are given by

\begin{equation}
    G'(\omega) = \frac{G_{0}(\eta_{0}\omega/G_{0})^{2} - \eta_{s}(G_{0}/\eta_{0})(\eta_{0}\omega/G_{0}^{2})}{(\eta_{0}\omega/G_{0})^{2}+1}
\label{storage}
\end{equation}

\begin{equation}
    G''(\omega) = \frac{\eta_{0} \omega + \eta_{s}\omega(\eta_{0}\omega/G_{0}^{2})}{(\eta_{0} \omega/G_{0}^{2}) + 1}
\label{loss}
\end{equation}

The experimental data on apparent viscosity, storage and loss moduli of Raghavan and Kaler~\citep{raghavan2001highly} have been fitted with the equations~\ref{app_vis},~\ref{storage}, and~\ref{loss}, as shown in figure~\ref{fitting}, and the corresponding MBM model parameters obtained and used in this study are as follows:    
$\lambda = 0.323 \,s$, $G_0 = 50.02 \,Pa$, $\frac{K_0}{\eta_\infty} = 0.323$, $\eta_{s} = 0.001 \,Pa \cdot s$, and $\eta_{0} = 6.956\,Pa \cdot s$.

\bibliographystyle{jfm}
\bibliography{jfm}
\end{document}